\documentclass[prl,singlecolumn,superscriptaddress,preprintnumbers,showpacs,floatfix,amsmath,amssymb]{revtex4-1}
\usepackage{epsfig}
\usepackage{graphics}
\usepackage{subcaption} 
\usepackage{float} 
\usepackage{array}
\usepackage{multirow}

\usepackage{yfonts}

\usepackage{graphicx}
\usepackage{amssymb}
\usepackage{mathrsfs}
\usepackage{bbm}
\usepackage{epsfig}
\usepackage{makecell}

\usepackage[usenames,dvipsnames]{color}

\usepackage{bbm}
\usepackage{color}
\usepackage{wrapfig}

\newcommand{\be}{\begin{eqnarray}}
\newcommand{\ee}{\end{eqnarray}}

\begin{document}
$~$
\vskip 3.0cm
\title{
\Large 
Time crystals: From Schr\"odinger to Sisyphus \\ ~ \\ } 
%
\author{Antti J. Niemi}
\email{Antti.Niemi@su.se}
\affiliation{
Nordita, Royal Institute of Technology, Stockholm University and Uppsala University, Hannes Alfv\'ens v\"ag 12, SE-106 91 Stockholm, Sweden}
\affiliation{
Pacific Quantum Center, Far Eastern Federal University,
				690950 Sukhanova 8, Vladivostok, Russia}
\affiliation{School of Physics, Beijing Institute of Technology, Haidian District, Beijing 100081, People's Republic of China}
%
\begin{abstract}\vskip 2.0cm
A Hamiltonian time crystal can emerge when a Noether symmetry is subject to a condition that
prevents the energy minimum from being a critical point of the Hamiltonian.  
A somewhat trivial example is the Schr\"odinger  equation of a harmonic oscillator. The Noether charge for its
particle number coincides with the square norm of the wave function, and the energy eigenvalue is a  Lagrange 
multiplier for the condition that  the wave function is properly normalized. A more elaborate
example is the Gross-Pitaevskii equation that models vortices in a cold atom Bose-Einstein condensate. 
In an oblate, essentially two dimensional harmonic trap the energy  minimum is a topologically protected
 timecrystalline vortex that rotates around the trap center. 
Additional examples are constructed using coarse grained Hamiltonian  models of 
closed molecular chains. When knotted, the topology of a chain can support a 
time crystal. As a physical example, high precision all-atom molecular dynamics is used to analyze an isolated 
cyclopropane molecule. The simulation reveals  that the molecular D$_{3h}$ symmetry 
becomes spontaneously broken.  When the molecule is observed with sufficiently long 
stroboscopic time steps it appears to rotate like a 
simple Hamiltonian time crystal. 
When  the length of the stroboscopic time step is decreased 
the rotational motion  becomes increasingly ratcheting and eventually
it resembles the back-and-forth oscillations of Sisyphus dynamics. The stroboscopic rotation is entirely due
to atomic level oscillatory shape changes,  so that  cyclopropane is an example of a molecule that 
can rotate without angular 
momentum. Finally, the article is concluded with a personal recollection how Frank's 
and Betsy's Stockholm journey started.  
\vskip 4.0cm
Based on talk at Frankfest, H\"ogberga 2021
\end{abstract}
\maketitle

\section{1. Introduction}\label{ra_sec1}

Wilczek ~\cite{wilczek-2012} together with Shapere  ~\cite{shapere-2012b} envisioned  a {\it  time crystal} to be a 
state that extends  the notion of spontaneous breakdown of space translation symmetry  to include 
time translation symmetry.  They argued that if time translation symmetry becomes spontaneously broken, 
the lowest energy state of a physical system can no longer be time independent but has to change with time. 
If the symmetry breaking leaves behind a discrete group the ensuing time evolution will be periodic, and in analogy 
with space crystals we have a time crystal. 

The  proposal drew immediate criticism: 
In a canonical Hamiltonian  setting the generator of time translations is the Hamiltonian itself.  
Thus energy is a conserved charge.  But if time translation symmetry becomes spontaneously 
broken,  the ground state energy can no longer be a  conserved quantity. 
Accordingly,  a time crystal must be impossible in any kind of isolated and energy 
conserving physical system, governed  by autonomous Hamiltonian
equation of motion  ~\cite{bruno-2013,watabane-2015,yamamoto-2015}.

As a consequence  the search of spontaneously broken time translation symmetry focused on driven,
non-equilibrium quantum systems 
~\cite{Sacha-2015,Sacha-2016,khemani-2016,else-2016a,else-2016,else-2016b,Yao-2017}.
The starting point is a many-body system with intrinsic dynamics governed by a
time independent Hamiltonian $H$. The system is then subjected to an external explicitly time periodic driving 
force, with period $T$, that derives from an extrinsic time dependent potential $U(t+T)= U(t)$. The total Hamiltonian 
is  the sum of the two $H_{tot}(t)=H+U(t)$ so that  the total Hamiltonian is also explicitly  time dependent,
with the same period $T$ of the drive.  Floquet theory  asserts that there can be solutions to the corresponding time 
dependent Schr\"odinger equation, with a time period that is different from $T$ where the difference is due to a  
Floquet index.  As a consequence it is plausible  that in such a periodically driven  
many-body system, a spontaneous self-organization of time periodicity takes place so that 
the system starts evolving with its own characteristic period, which is different from that of the 
external driving force. It was first shown numerically that this kind of self-organisation can take 
place in certain many-body localized spin systems ~\cite{Sacha-2015,Sacha-2016,khemani-2016,else-2016a,else-2016,else-2016b,Yao-2017}.
Experiments were then
performed in appropriate material realizations  ~\cite{zhang-2017,lukin-2017,Rovny-2018} and they confirmed
the presence of
sustained collective oscillations with a time period that  is indeed different from the 
period of the external driving force. It is now widely accepted that  this kind of driven non-equilibrium 
Floquet time crystals do exist, but the setting  is quite distant from Wilczek's original idea.

Here I  describe how, in spite of the {\it No-Go} arguments, {\it genuine}  (semi)classical and 
quantum Hamiltonian time crystals do exist~\cite{Dai-2019,anton,Dai-2020}. They could even be widespread.  I start with an  
explicit {\it proof-of-concept} construction that shows how to go around the {\it No-Go} arguments: I show 
that whenever a Hamiltonian system supports conserved Noether charges that
are subject to appropriate conditions,  
the  lowest energy ground state can be both time dependent and have an energy value that is 
conserved. This can occur whenever the lowest energy ground state, as a consequence of the conditions, 
is not a critical point of the 
Hamiltonian. I explain how such a time dependent ground state can  be explicitly 
constructed by the methods of constrained optimization~\cite{Fletcher-1987,Nocedal-1999}, using the 
Lagrange multiplier theorem~\cite{Marsden}.
Whenever the solution for {\it any}  Lagrange multiplier is non-vanishing, the lowest energy ground state is time
dependent:  A time crystal is simply a time dependent symmetry transformation that
converts the time evolution of a Hamiltonian system into  an equivariant time evolution. 
In particular, since time translation symmetry is not spontaneously broken but 
equivariantized, unlike in the case of conventional spontaneous symmetry breaking now
there is no massless Goldstone boson that can be associated with a time crystal. But the 
two concepts can become related in a limit where the 
period of a time crystal goes to infinity and its energy approaches a (degenerate) critical point of the Hamiltonian.


I present a number of examples, starting with  the time dependent Schr\"odinger equation of a harmonic oscillator. I 
continue with the Gross-Pitaevskii equation that
models vortices in a cold atom Bose-Einstein condensate, and with a generalized  Landau-Lifschitz equation that 
can describe closed molecular chains in a coarse grained approximation. I then analyze cyclopropane as 
an actual molecular example of 
timecrystalline Hamiltonian dynamics. For this I employ high precision all-atom molecular dynamics to investigate the 
ground state properties of  a single isolated cyclopropane molecule.  I conclude that the maximally symmetric configuration,
with the D$_{3h}$ molecular symmetry, can become spontaneously broken. I follow the time evolution of the minimum
energy configuration stroboscopically, at very low but constant internal  temperature values. 
I find that with a proper internal temperature and  for sufficiently long stroboscopic time steps the molecule becomes a
Hamiltonian time crystal with time evolution that is described by the generalization of the Landau-Lifschitz equation. 
But when the length of the  stroboscopic time step is decreased, there  is a cross-over transition to an increasingly  ratcheting
time evolution. In the limit where the stroboscopic time 
step is very small the motion of the cyclopropane molecule resembles Sisyphus dynamics ~\cite{sisyphus}.
I propose that this kind of cross-over transition  between the Sisyphus dynamics that describes the limit of very short
stroboscopic time steps, and the uniform 
timecrystalline  dynamics that describes the limit of very long stroboscopic time steps, is  
a universal phenomenon. It exemplifies that the coarse graining of an apparently random 
microscopic many-body system can lead to a separation of scales and
self-organization towards an effective theory Hamiltonian time crystal.   

Finally, the  rotational motion that I observe in a cyclopropane in the limit of long
stroboscopic time steps, can occur even with no angular momentum.
Thus, cyclopropane is a molecular level example of a general phenomenon of rotation by shape deformation, and 
without any angular momentum: 
The short time scale vibrational motions of 
the individual atoms that are driven {\it e.g.} by thermal or maybe even by quantum mechanical zero point fluctuations 
can become converted into a large scale rotational motion of the entire molecule, even when there is no 
angular momentum. This kind of phenomenon was first predicted  by Guichardet~\cite{Guichardet-1984}, and independently 
by Shapere and Wilczek~\cite{Shapere-1989b}, and an early review can be found in ~\cite{Littlejohn}.

%
%
%
%
%
%
%
%
%
%
%
%
%

\section{2. Hamiltonian time crystals}\label{sect-1}

A Hamiltonian time crystal describes the minimum of a Hamiltonian that is not a critical point of the Hamiltonian.
To show how this can occur,  I start with  the Hamiltonian  action
\begin{equation}
S = \int dt \ p_i \dot q_i - H(p,q)
\label{S}
\end{equation}
where ($p_i, q_i$) ($i,j=1,...,N$) are the local Darboux coordinates with non-vanishing Poisson bracket
\begin{equation}
\{ q_i, p_j\} = \delta_{ij}
\label{PB}
\end{equation}
and Hamilton's equation is
\begin{equation}
\begin{split}
& \frac{dq_i}{dt} = \{ q_i , H \} = \ \ \frac{\partial H}{\partial p_i}
\\
& \frac{dp_i}{dt} = \{ p_i , H \} =  - \frac{\partial H}{\partial q_i}
\end{split}
\label{pqH}
\end{equation}
On a compact closed manifold the minimum of $H(p,q)$ is also its critical point, and  in that case the
Hamiltonian can not support any time crystal~\cite{bruno-2013,watabane-2015,yamamoto-2015}. 
Thus for a time crystal we need additional structure, and for this
I focus on a canonical   Hamiltonian system that is subject to appropriate conditions
\begin{equation}
G_a(p,q) - g_a = 0 \ \ \ \ \ \ \  \ \ \ \ \ a=1,...,n \leq  N
\label{G}
\end{equation}
where the $g_a$ are some constants and  the $G_a$'s define a Noether symmetry,
\begin{equation}
\begin{split}
\{ G_a , G_b  \} = &  \ f_{abc} G_c  \ \ \ \ \ \ {\rm for \ \  all } \ \ a,b,c \\
\{ G_a , H \}  =  & \ 0 
\end{split}
\label{symm}
\end{equation}
I  then search for a solution to the following problem~\cite{anton}:

\vskip 0.2cm
\noindent
\hskip 2.0cm {\it First, find  a minimum of $H(p,q)$ that is subject to appropriate conditions  (\ref{G}) but is  not a 

\hskip 1.6cm critical point of $H(p,q)$. Then,  solve (\ref{pqH}) with this  minimum as the initial condition. }
\vskip 0.2cm

\noindent
The first step  is a classic problem in constrained optimization~\cite{Fletcher-1987,Nocedal-1999}
and it can be solved using the Lagrange multiplier theorem~\cite{Marsden}. 
The  theorem states
that the minimum of $H$ can be found as a critical point of
\begin{equation}
H_\lambda(p,q;\lambda) \ = \ H(p,q) + \lambda_a (G_a(p,q) - g_a)
\label{Hlambda}
\end{equation}
where the $\lambda_a$ are independent,  {\it a priori} time dependent auxiliary variables.
The critical point of (\ref{Hlambda}) is a solution of
\begin{equation}
\begin{split}
& \frac{ \partial H }{\partial  p_i} + \lambda_a \frac{ \partial G_a}{\partial  p_i} \ = \ 0 
\\
&  \frac{ \partial H }{\partial  q_i} + \lambda_a \frac{ \partial G_a}{\partial  q_i} \ = \ 0 
\\ 
& 
G_a(p,q) \  = \  g_a
\end{split}
\label{eqslambda}
\end{equation}
Since the number of equations (\ref{eqslambda}) equals the number  of unknowns ($p,q,\lambda$) a 
solution ($p^\star, q^\star, \lambda^\star$) including the Lagrange multiplier, can be found at least in principle.
Under proper conditions, in particular if  $H(p,q)$ is strictly convex and the $G^a(p,q)$ are affine functions, existence and
uniqueness theorems can also be derived.  But if there are  more than one solution  I choose the one with
the smallest value of $H(p,q)$.

Suppose the solutions  $\lambda^\star_a$ for the Lagrange multipliers do not all vanish.  By combining
(\ref{eqslambda}) with (\ref{pqH}) I conclude that I have a time crystal~\cite{anton} 
with the initial condition
\begin{equation}
p_i(t=0) = p_i^\star \ \ \ \ \& \ \ \ \ q_i(t=0) = q_i^\star \ \ \ \ \& \ \ \ \   \lambda_a(t=0) =  \lambda_a^\star
\label{ini}
\end{equation}
and with the time evolution
\begin{equation}
\begin{split}
& \frac{dq_i}{dt} \ = \ -  \lambda_a^\star \frac{ \partial G_a}{ \partial p_i }  \ = \  -  \lambda_a^\star \{ q_i , G_a  \} 
\\
& \frac{dp_i}{dt} \ =  \   \   \ \lambda_a^\star \frac{ \partial G_a}{ \partial q_i }  \  = \  -  \lambda_a^\star \{ p_i , G_a  \}
\end{split}
\label{pqstar}
\end{equation}
and the  Lagrange multipliers can be shown to be time independent $ \lambda_a(t) \equiv \lambda_a^\star$. 
In particular, the timecrystalline evolution (\ref{pqstar})  determines  
a time dependent symmetry transformation of the minimum energy configuration (\ref{ini}), one that is generated by the linear combination
\[
G^\lambda_a \ = \ \lambda_a^\star G_a
\]
of the Noether charges. 

Since the Hamiltonian $H$ has no explicit time dependence
I immediately conclude that the  energy of the time crystal is a conserved quantity
\[
\frac{dH}{dt} \ = \ -  \lambda_a^\star \{ H ,  G_a \} \ = \ 0
\]
This contrasts some of the early arguments, to exclude a time crystal on the grounds that the minimum energy should be time dependent. 

In the sequel I consider exclusively such energy conserving Hamiltonian time crystals, with a time evolution that is a symmetry transformation.
%
%
%
%
%
%
%
%
%
%

\section{3. Schr\"odinger as time crystal }\label{sect-2}

For a simple example of the general formalism, I start with the following canonical action
\begin{equation}
S = \int dt d\mathbf x \ \{ \ \bar \psi (i \partial_t) \psi -  \bar\psi (- {\nabla}^2 + |\mathbf x|^2 )\psi \ \}  
\label{schact}
\end{equation}
in $D$ space dimensions. This action yields the non-vanishing Poisson bracket
\begin{equation}
\{ \bar \psi (\mathbf x_1) , \psi (\mathbf x_2) \} = - i \delta(\mathbf x_1 - \mathbf x_2 )
\label{schbr}
\end{equation}
The  Hamiltonian energy is
\begin{equation}
H \ = \ \int d\mathbf x \, \bar\psi (-\nabla^2 + |\mathbf x|^2 ) \psi 
\label{SchH}
\end{equation}
and Hamilton's equation coincides with the time dependent Schr\"odinger equation
\begin{equation}
i\partial_t \psi = - \nabla^2 \psi + |\mathbf x|^2 \psi
\label{schrt}
\end{equation}
The Hamiltonian  is strictly convex and its unique critical point  is the absolute minimum 
\[
\psi(\mathbf x ) \ = \ 0
\]
At this point I have an example  of the situation governed by (\ref{pqH}). In particular,  there is no time crystal, as defined
in the previous Section.

To obtain a time crystal I follow the general formalism of the previous section: I introduce  the square norm
\begin{equation}
N =   \int d\mathbf x \ \bar\psi \psi
\label{N-ch}
\end{equation}
This is the Noether charge  for the symmetry of (\ref{schact}) under a phase rotation, it counts the number of particles,
and I subject it to the following familiar condition (\ref{G}),
\begin{equation} 
G_1 \ \equiv  \ N-1  =   (\int d\mathbf x \ \bar\psi \psi)  \ - 1  \ = \ 0
\label {N}
\end{equation}
I then  proceed with the Lagrange multiplier theorem and search for the critical point of the corresponding functional 
(\ref{Hlambda})
\[
H_E = \int d\mathbf x \, \bar\psi (-\nabla^2 + |\mathbf x|^2 ) \psi  \ - \ E ( \int d\mathbf x \ \bar\psi \psi  \ - 1 )
\]
 Where $E \equiv -\lambda $ is the Lagrange multiplier:  
The corresponding equations (\ref{eqslambda}) coincide with the time independent Schr\"odinger equation for a harmonic
oscillator
\[ 
\begin{split}
& -\nabla^2 \psi + |\mathbf x|^2 \psi = E\psi \\
& \int d\mathbf x \ \bar\psi \psi  = 1  
\end{split}
\]
In general there are many solutions,  all the harmonic oscillator eigenstates are solutions.    
But I pick up the solution ($\psi_{min}^\star(\mathbf x) , E_{min}^\star$) 
that minimizes the energy (\ref{SchH}). This is exactly the textbook lowest energy 
ground state wave function 
$\psi_{min}^\star(\mathbf x)$ of the $D$ dimensional harmonic oscillator, and $E_{min}^\star$  is the corresponding  
lowest energy eigenvalue.  

In line with (\ref{ini}), (\ref{pqstar}) I can write the time dependent Schr\"odinger equation (\ref{schrt}) as follows,
\begin{equation}
i\partial_t \psi \ = \  E_{min}^\star \{ N , \psi \} \ \equiv \ E_{min}^\star \{ \int \bar\psi \psi , \psi \} \ \ \ \ \ \ {\rm with} \ \ \ \ \ \ \psi(\mathbf x, t=0) \ = \  \psi_{min}^\star(\mathbf x)
\label{schtime}
\end{equation}
That is, the time evolution of the harmonic oscillator wave function is a symmetry transformation generated by the Noether charge $N$
{\it i.e.} a phase rotation, with the familiar solution 
\begin{equation}
\psi(\mathbf x,t) = e^{-iE_{min}^\star t}\, \psi_{min}^\star(\mathbf x)
\label{solusch}
\end{equation}
Normally, a time dependent phase factor is not an observable. But it can be made so, {\it e.g.} in a proper double slit experiment. 
Note  that unlike in the case of standard spontaneous symmetry breaking, 
even though time translation invariance is converted  into an {\it equivariant}  time translation, 
there is no Goldstone boson in a quantum harmonic oscillator.

The previous example is   {\it verbatim}  a realization of the general formalism in Section 2.  Albeit  quite
elementary in its familiarity and simplicity, it nevertheless makes the point. Conditions such as (\ref{G}) on Noether 
charges are commonplace, they often have a pivotal role in specifying the physical scenario. When that happens, a time
crystal can  appear.  Moreover, without additional structure the appearance of a time crystal 
does not entail the emergence of a massless Goldstone boson:
A time crystal does not break the time translation symmetry. Instead, it  equivariantizes a time translation 
into a combination of  a time translation and
a symmetry transformation.

%
%
%
%
%
%
%
%
%
%

\section{4. Nonlinear Schr\"odinger and  time crystalline vortices }\label{sect-3}

I now proceed with additional examples of the general formalism. For this 
I observe that besides phase rotations generated by the Noether charge $N$, 
the Schr\"odinger action (\ref{schact}) has  also a Noether symmetry under space rotations. 
Accordingly I introduce the
corresponding Noether charges 
\[
\mathbf L  \ = \ \int d\mathbf x \, \bar\psi ( - i \mathbf x \wedge \nabla )\psi \,
\]
Their Poisson brackets coincide with the Lie algebra $SO(D)$ in $D$ dimensions, and 
have vanishing Poisson brackets  with the conserved charge (\ref{N-ch}),
\[
\{ \mathbf L , N \} = 0
\]
I follow the general formalism: I impose  additional conditions to the harmonic oscillator using  
the maximal commuting subalgebra of space rotations;  in three dimensions this 
amount to the familiar conditions  $\mathbf  L^2=l(l+1)$ and  $L_z=m$ with $l\in \mathbb Z^+$ and $m 
\in \mathbb Z$ and $|m|\leq l$.  I can take results from any quantum mechanics textbook and 
confirm that all the higher angular momentum 
eigenstates of the harmonic oscillator yield the appropriate minimum energy solutions of the 
time dependent  Schr\"odinger equation.  

For a more elaborate structure,  I  introduce a quartic self-interaction and consider the following
nonlinear Schr\"odinger action, 
\begin{equation}
S = \int dt d\mathbf x \ \{ \ \bar \psi (i \partial_t) \psi -  \bar\psi (- \nabla^2 + |\mathbf x|^2 )\psi - \frac{g}{2} (\bar \psi \psi)^2\ \}  
\label{schact2}
\end{equation}
This action also supports both $N$ and $\mathbf L$ as conserved  Noether charges. 
It  defines the 
Gross-Pitaevskii model that describes {\it e.g.} the Bose-Einstein condensation of cold alkali atoms at the level of a 
mean field theory~\cite{sonin}, with $g>0$ the quartic nonlinearity models short distance two-body repulsion. 
For clarity I specify to two space dimensions so that there is only one conserved charge $L\equiv L_z$.
The set-up is that of a spheroidal, highly oblate three dimensional trap,  with atoms  
confined to a disk by a very strong trap potential in the $z$-direction, and much weaker harmonic trap  potential in 
the ($x,y$) direction.  Thus,  besides the condition (\ref{N}) I  also introduce 
the following  condition
\begin{equation}
G_2 \ \equiv \ L_z - l_z \ = \ \int dxdy \ \bar\psi  (-i \mathbf x \times \nabla ) \psi \ - \ l_z \ = \ 0
\label{G2}
\end{equation}
so that I have the scenario  (\ref{G}), (\ref{symm}) with two Noether charges. 

I keep the normalization (\ref{N}) but I 
leave the numerical value $l_z$ of the angular momentum 
as a free parameter.
The choice  (\ref{N}) is entirely for convenience:  I can always rescale the square norm to set $N$ to 
any non-vanishing  value, and compensate for this by 
adjusting  the length and time scales. But  $l_z$  and the
parameter $g$ in (\ref{schact2}) remain as freely adjustable parameters~\cite{Garaud-2021a}.
Note that even though the microscopic, individual atom angular momentum is certainly quantized, 
the macroscopic  angular momentum $L_z$  does not need to be quantized: In applications to cold atoms the 
wave function $\psi(x,y)$ is a macroscopic condensate wave 
function that describes a very large number of atoms. Thus, it can support arbitrary values of the 
macroscopic angular momentum $L_z$.

Following  \cite{Garaud-2021a} I search for a  solution of the nonlinear time dependent
Schr\"odinger equation of (\ref{schact2}) {\it a.k.a.}  the Gross-Pitaevskii equation
\begin{equation}
i\partial_t \psi = -\nabla^2 \psi + |\mathbf x|^2 \psi + g |\psi |^2 \psi 
\label{nlse}
\end{equation}
that minimizes the Hamiltonian {\it i.e.} energy
\begin{equation}
H = \int d\mathbf x \  \{ \bar\psi (- \nabla^2 + |\mathbf x|^2 )\psi + \frac{g}{2} (\bar \psi \psi)^2\}
\label{F}
\end{equation}
subject to the two conditions (\ref{N}) and (\ref{G2}). 
The Hamiltonian is strictly convex, its global minimum coincides with the only critical point $\psi(\mathbf x)\equiv 0$. 
But when it  is subject  to  the conditions (\ref{N}) and (\ref{G2}) the minimum does not longer need to be
a critical point. Instead, the Lagrange multiplier theorem 
states that the minimum is  a critical point ($\psi^\star, \lambda_1^\star, \lambda_2^\star$) of
\begin{equation} 
H_\lambda \ = \ H + \lambda_1 G_1 + \lambda_2 G_2
\label{Flambda}
\end{equation}
That is, the minimum is a solution of
\begin{equation}
\begin{split}
& \frac{\delta H_\lambda}{\delta \psi} \ = \   -\nabla^2 \psi + |\mathbf x|^2 \psi + g |\psi |^2 \psi +  \lambda_1 \psi + \lambda_2 (-i\mathbf x \times \nabla ) \psi = 0 
\\
& \frac{\delta H_\lambda}{\delta \lambda_1} \ = \   \int d\mathbf x \ \bar\psi\psi -1 = 0
\\
&  \frac{\delta H_\lambda}{\delta \lambda_2} \ = \ \int d\mathbf x \ \bar\psi ( -i \mathbf x \times \nabla ) \psi \ - l_z = 0
\end{split}
\label{nlseF}
\end{equation}
If there are several solutions, I  choose ($\psi^\star, \lambda_1^\star, \lambda_2^\star$) 
for which  $H(\psi)$ has the smallest value. The minimum energy 
$\psi^\star(\mathbf x)$ then defines the initial condition
\begin{equation}
\psi(\mathbf x, t=0) = \psi^\star(\mathbf x)
\label{psiini}
\end{equation}
 for the putative timecrystalline solution of (\ref{nlse}); recall that the corrresponding  Lagrange multipliers
 are space-time independent, $\lambda_{1,2}(\mathbf x,t) \equiv  \lambda_{1,2}^\star$.
 
Remarkably, when I  combine (\ref{nlseF}) and  (\ref{nlse}),  
the {\it nonlinear} time evolution equation of $\psi^\star(\mathbf x)$ becomes converted to the following 
{\it linear} time evolution equation
\begin{equation}
 i\partial_t \psi \  = \   - \lambda_1^\star \psi +  i \lambda_2^\star  \mathbf x \times \nabla  \psi 
\  \equiv \ - \lambda_1^\star \{ G_1, \psi\} - \lambda_2^\star \{ G_2, \psi\}
\label{nlset}
\end{equation}
This is the equation (\ref{pqstar}) in the present case. In particular,  (\ref{nlset}) states that 
the time evolution of $\psi^\star(\mathbf x)$  is a symmetry transformation; the pertinent symmetry is a combination
of phase rotation by $N$ and a spatial rotation by $L_z$. 

Since the Hamiltonian $H$  
has vanishing Poisson brackets with both $G_1$ and $G_2$,  the energy is conserved during the timecrystalline evolution,
\[
\frac{d}{dt}  H  \ = \  \lambda_1^\star \{ H,G_1 \}  +  \lambda_2^\star \{ H,G_2 \} =0
\]

The solutions of (\ref{nlseF}), (\ref{nlse}) describe vortices that rotate uniformly, with angular velocity $-\lambda_2^\star$ 
around of the trap center. For this I define
\[
\mathbf  L_z = -i\mathbf x \times \nabla \ \ \ \ \ \ \& \ \ \ \ \ \ \mathbf A = \nabla \tan^{-1} \left(\frac{x}{y}\right)
\]
to write (\ref{nlset}) as follows:
\begin{equation}
i \partial_t \psi = -\lambda_2^\star \left(\mathbf  L_z  +  \frac{ \lambda_1^\star}{\lambda_2^\star}  \mathbf x \times \mathbf A \right) \psi \ \equiv \ \omega \mathbf L_z^{cov} \psi
\label{rota}
\end{equation}
Here $\mathbf L_z^{cov} $ is the covariant angular momentum that generates the  
rotations around the trap center in the presence of the "analog gauge field" $\mathbf A$. Note that $\mathbf A$ 
supports a "magnetic" flux with "strength" $\lambda_1^\star / \lambda_2^\star$ along the $z$-axis, that can have
non-integer values.  Notably,  $\mathbf L_z^{cov} $  has the same form as the angular momentum operator introduced
by Wilczek~\cite{Wilczek-1982a,Wilczek-1982b}, in the case of an anyon pair,  in terms  of the 
relative coordinate.

\begin{figure}
\centerline{\includegraphics[width=13.0cm]{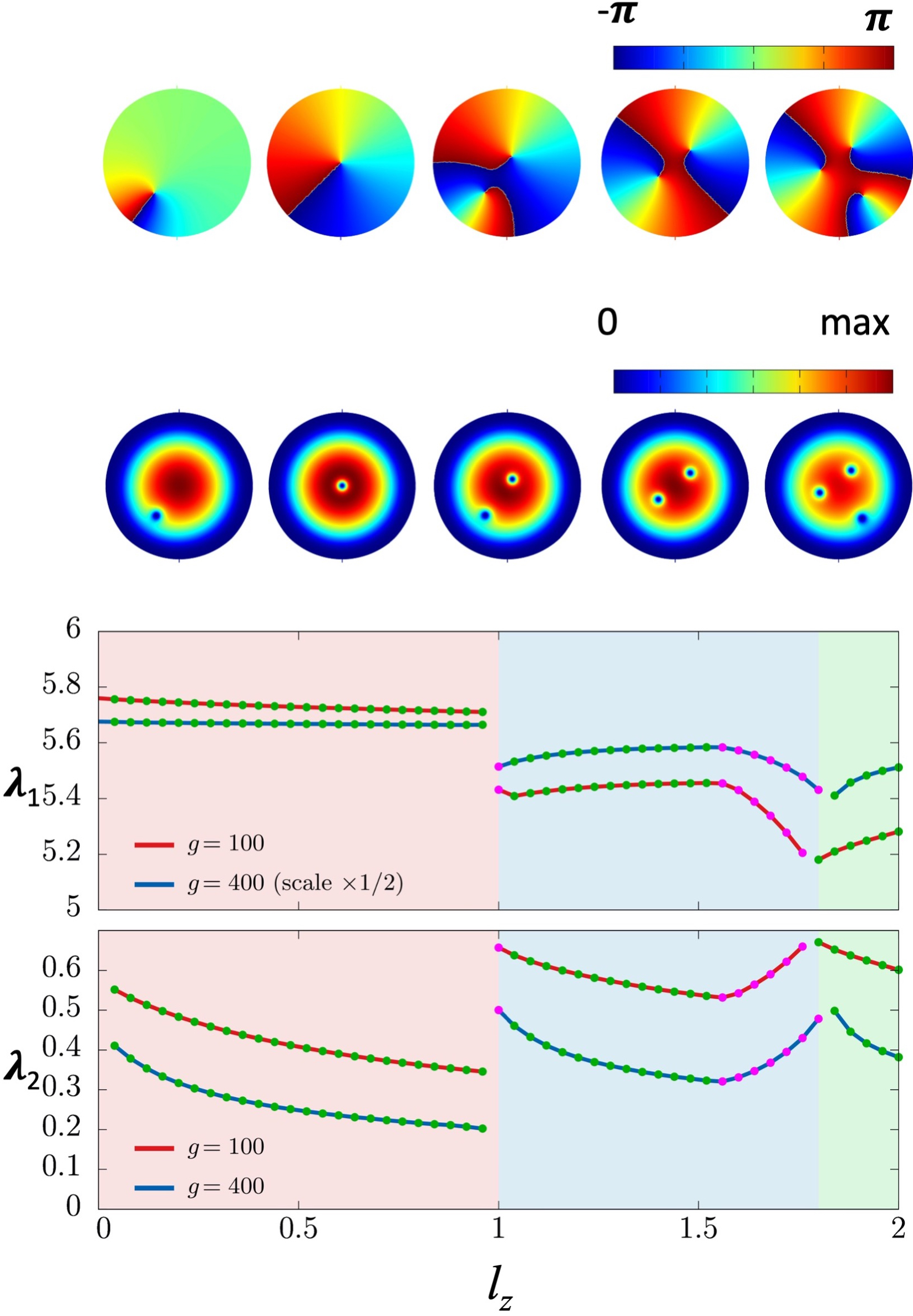}}
\caption{The first line shows the phase of the wave function $\psi^\star(\mathbf x,t)$. The second line shows the 
density $|\psi^\star(\mathbf x,t)|$ for five representative 
instantaneous minimum energy
vortex solutions of (\ref{nlse}), (\ref{psiini}): From left to right $0<l_z < 1.0$, $l_z=1$, $0<l_z  < 1.5$, $1.5 < l_z < 1.8$, $1.8 < l_z < 2.0$; the
numerical values $1.5, \ 1.8$ are approximative, and depend on the coupling $g$. The vortices rotate around the trap center with angular 
velocity determined by $\lambda_2^\star$.
In these figures the value of coupling is $g=400$ which is representative for cold atom Bose-Einstein condensates ~\cite{Garaud-2021a}.
The bottom panels show the evolution of Lagrange multipliers $\lambda_{1,2}^\star$ for $g=100$ and $g=400$.}
\label{fig-1}
\end{figure}

The Figure \ref{fig-1} summarizes the results from the numerical investigations  in [~\c{Garaud-2021a}].
The number of vortices and their relative positions including the 
distances from the trap center,  depend on the free parameter $l_z$.  
The two top lines of Figure \ref{fig-1} show how the phase and the cores  of the vortex structure evolve for $0<l_z<2$.   
The bottom panels show the corresponding values of the Lagrange
multipliers $\lambda_1^\star$ and $ \lambda_2^\star$.
%
%
%
%
%
%
%
%
%
%
%
Remarkably, when $l_z\to 0$ neither of these multipliers  vanish,
\begin{equation}
\lim_{l_z \to 0}  \lambda_{1,2}^\star(l_z) \not=0
\label{lim}
\end{equation}
Since $ \lambda_2^\star$ determines the angular velocity around the trap center (\ref{rota}), this means 
that for any $l_z\not=0,1$ the minimum energy solution describes a rotating vortex configuration. 

For $l_z=1$ the vortex core coincides with the trap center, only its phase has time dependence.  For $l_z=0$ 
the   minimum energy solution is a real valued function, up to a constant phase factor.  
For  $0<|l_z|<1$ the distance between 
the vortex  core, located at point ${\tt p}\in \mathbb R^2$, and the trap center  increases as $|l_z|$ decreases. 
Notably, the limit $l_z\to 0$ is discontinuous and to inspect this I  introduce the superfluid velocity
\begin{equation}
\mathbf v (\mathbf x,t) = \nabla \arg[\psi] (\mathbf x,t)
\label{v}
\end{equation}
and I define its  integer valued circulation 
\begin{equation}
i_{\mathbf v}({\tt p};\Gamma) \ = \ \frac{1}{2\pi} \oint_\Gamma d{\boldsymbol \ell}  \cdot \mathbf v
\label{circu}
\end{equation}
where $\Gamma$ is a closed trajectory on the plane that does not go thru any singular point of $\mathbf v (\mathbf x,t)$.
For any given $l_z \not=0$ I can always choose $\Gamma$ to be 
a circle around the trap center and with a large enough radius, so that the core  ${\tt p}$ of a given  
vortex
is inside this circle. For a single vortex the value of  (\ref{circu}) is $i_{\mathbf v}({\tt p};\Gamma) = \pm 1$, with positive sign 
for $l_z>0$ and negative for $l_z<0$. 
For $l_z=0$ the value of the integral (\ref{circu})  vanishes, and for this value  
the phase of $\psi(\mathbf x,t)$ can be chosen to vanish;
the entire plane is a fixed point of (\ref{v}) for $l_z=0$.  

The  circulation (\ref{circu})  is an integer valued  topological invariant. In particular it can not be continuously deformed as a function of
 $l_z$,  when $l_z$ varies between $i_{\mathbf v}({\tt p};\Gamma)=+1$ for $l_z>0$ and $i_{\mathbf v}({\tt p};\Gamma)=-1$ for $l_z<0$.
When  $|l_z|$ continues to increase to  values $|l_z|>1$ 
the value of  (\ref{circu}) increases but always in integer steps, as new vortex cores enter 
inside the (sufficiently large radius) circle $\Gamma$.
Thus the  vortex structures are topologically protected timecrystalline solutions of (\ref{nlse}).

%
%
%
%
%
%
%
%
%
%

\section{5. Timecrystalline molecular chains}\label{ra_sec4}

I now proceed to  a general class of Hamiltonian time crystals. The
Hamiltonians describe discrete, piecewise linear chains~\cite{Dai-2019,Dai-2020}.  
One  can think of the vertices $\mathbf r_i $ ($i=1,...,N$) as point-like interaction centers, they  can {\it e.g.} model atoms or small 
molecules in a coarse grained description of a linear polymer.  
The links $\mathbf n_i = \mathbf r_{i+1} - \mathbf r_i$ between the vertices then model  {\it e.g.}
the covalent  bonds, or peptide planes in the case of a protein chain. 
For clarity I only consider cyclic chains, with the convention 
$\mathbf r_{N+1} = \mathbf r_1$. 

In an actual molecule, the covalent bonds are very rigid and oscillate rapidly, with a characteristic period as short as 
a few femtoseconds. 
I am  mostly interested in timecrystalline dynamics at much longer time scales. Thus I assume that  the lengths 
of the links are constant, and equal to their time averaged values in  the underlying atomistic system. For simplicity
I take all the links to have an equal length with the numerical value 
\begin{equation}
|\mathbf n_i |  \equiv  |\mathbf r_{i+1} - \mathbf r_i| = 1
\label{nr}
\end{equation}
%
%
%
%
%
%
\begin{figure}
\centerline{\includegraphics[width=15.0cm]{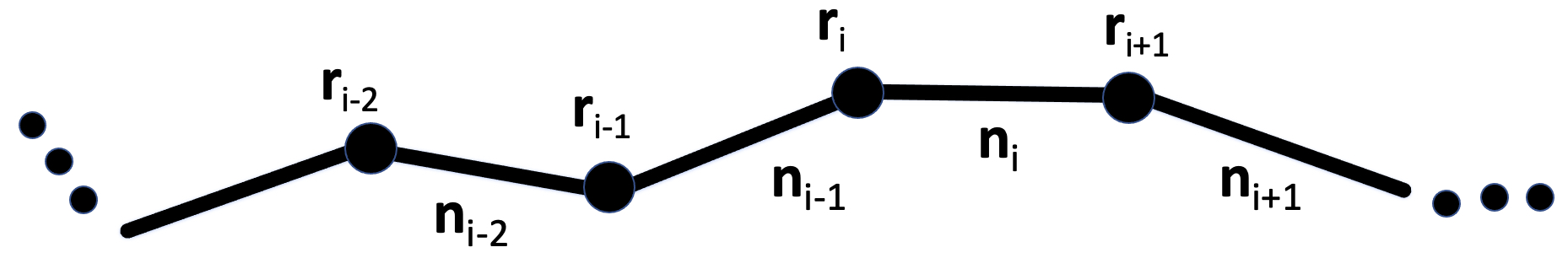}}
\caption{ The dynamical variables $\mathbf n_i  =  \mathbf r_{i+1} - \mathbf r_i$ are links that connect the vertices $\mathbf r_i$ of
a piecewise linear chain.}
\label{fig-2}
\end{figure}
%
%
%
%
%
The  link variables $\mathbf n_i$ are the dynamical coordinates in my set-up. I subject them to the 
Lie-Poisson bracket~\cite{Dai-2019}
\begin{equation}
\{ n_i^a , n_j^b \} = \pm \epsilon^{abc} \delta_{ij} n^c_i 
\label{n-bra}
\end{equation}
where I can choose either sign on the {\it r.h.s.} and for clarity I choose $+$-sign; the two signs are related by parity {\it i.e.} change in direction of $\mathbf n_i$. 
The bracket is designed to generate all possible local motions of the chain except for stretching and shrinking of its links, 
\[
 \{ \mathbf n_i , \mathbf n_k \cdot \mathbf n_k \} =0 
 \] 
 for all pairs ($i,k$), so that (\ref{nr}) is preserved for all times,  independently of the 
Hamiltonian details. 

Remarkably,  the same 
bracket (\ref{n-bra})  appears in Kirchhoff's theory of a rigid body that moves in an unbounded incompressible 
and irrotational fluid that is at rest at infinity~\cite{Milne}.

Hamilton's equation coincides with  the Landau-Lifschitz equation 
\begin{equation}
\frac{ \partial \mathbf n_i}{\partial t} = \{  \mathbf n_i , H  \} 
=  \mathbf n_i \times  \frac{\partial H}{\partial \mathbf n_i}  
\label{eom}
\end{equation}
and the condition that the chain is closed gives rise to the first class constraints 
\begin{equation}
\begin{split}
& \mathbf G  \equiv  \sum_{i=1}^{N} \mathbf n_i  \ = \ 0 \\ 
& \{ G^a, G^b \} = \epsilon^{abc} G^c \\
& \{ \mathbf G , H \} = 0 
\end{split}
\label{first}
\end{equation}
The last equation restricts the form of Hamiltonian functions I consider.
Such Hamiltonians  $H_i(\mathbf n)$   
can be constructed {\it e.g.} as linear combinations of
the Hamiltonians that appear in the integrable hierarchy of the Heisenberg chain:
\begin{equation}
\begin{split}
H_1 = & \sum\limits_{i} a_i \mathbf n_i \cdot \mathbf n_{i+1} \\
H_2 = & \sum\limits_{i} b_i \mathbf n_i \cdot (\mathbf n_{i-1}\times \mathbf n_{i+1} ) \\
H_3 = & \sum\limits_i c_i \mathbf n_i \cdot ( \mathbf n_{i-1} \times (\mathbf n_{i+1} \times \mathbf n_{i+2} )) \\
 H_4 = & \sum\limits_{i} d_i \mathbf n_{i-1} \cdot \mathbf n_{i+1} \\
{\it etc.} &
\end{split}
\label{H1234}
\end{equation}
where $a_i,b_i,c_i,d_i$ are parameters. Furthermore, 
\begin{equation}
\mathbf r_i - \mathbf r_j  =  \frac{1}{2} ( \mathbf n_j + ... + \mathbf n_{i-1}  - \mathbf n_{i} - ... - \mathbf n_{j-1}) 
\label{rn}
\end{equation}
where I have symmetrized the distance, since there are two ways to connect $\mathbf r_i$ and  $ \mathbf r_j $ along the chain.
As a consequence I  can also introduce two-body interactions between distant vertices  as contributions 
in a Hamiltonian, 
such as a combination of 
electromagnetic Coulomb potential and the Lennard-Jones potential:  
\begin{equation}
U(\mathbf x_1,...,\mathbf x_N)  = \frac{1}{2} 
{\sum_{\buildrel{i,j=1}\over{i\not=j}}^N}  \frac{ e_i e_j  } { |\mathbf x_{i} - \mathbf x_j | }
+ \frac{\epsilon}{2} {\sum_{\buildrel{i,j=1}\over{i\not=j}}^{N}} \left\{   \left(\frac{ \sigma_{P} } { |\mathbf x_{i} - \mathbf x_j | }\right)^{\!\! 12} - \left(\frac{ \sigma_{vdW} } { |\mathbf x_{i} - \mathbf x_j | }\right)^{\!\! 6}\right\}
\label{U}
\end{equation} 
Here $e_i$ is the charge at the vertex $\mathbf x_i$,  $\sigma_{P} $ characterizes the extent of the Pauli exclusion 
that  prevents chain crossing, and $\sigma_{vdW} $ is the range of the van der Waals interaction.
All  are commonplace in molecular modeling, and in particular they comply with  the conditions (\ref{first}). 

Thus, the combination of (\ref{H1234}) and  (\ref{U}) can be employed to construct realistic
coarse grained  Hamiltonian functions $H$ that describe dynamics of (closed) molecular chains, in a way that 
only depends on the vectors $\mathbf n_i$. 

Note in particular that (\ref{first}) implies that 
an initially closed chain remains a closed chain during the 
time evolution (\ref{eom}).

I follow  the general formalism of Section 2 to search for a  time crystal as a 
minimum of $H(\mathbf n)$ that is subject to the chain closure condition (\ref{first}).
The minimum is a critical point of  the following version of (\ref{eqslambda}), (\ref{pqstar})
\begin{equation}
\begin{split}
& H_{\boldsymbol \lambda} = H + {\boldsymbol \lambda} \cdot \mathbf G \\
& \frac{\partial H}{\partial \mathbf n_i}_{| \mathbf n^\star} = - {\boldsymbol \lambda}^\star \ \ \ \ \ \ \  \& \ \ \ \ \ \mathbf G  (\mathbf n^\star)= 0 \\
\Rightarrow \ \ & \frac{\partial H}{\partial t} = \ {\boldsymbol \lambda}^\star \times \mathbf n_i \ \ \ \ \  \& \ \ \ \ \ \mathbf n_i(t=0) = \mathbf n_i^\star 
\label{HL2}
\end{split}
\end{equation}
Whenever the solution $\boldsymbol \lambda^\star \not=0$ I have a time crystal as a closed polygonal 
chain that rotates like a rigid body. The direction of its rotation and its angular velocity are both determined by the direction and the magnitude of the time independent vector $\boldsymbol \lambda^\star$. Thus, the present  timecrystalline
Hamiltonian framework provides an effective theory framework for modeling the autonomous dynamics of
rotating cyclic molecules.

In practice, to construct the minimum of $H(\mathbf n)$ 
I extend (\ref{eom}) to the Landau-Lifschitz-Gilbert equation for $H_{\boldsymbol \lambda}$~\cite{anton,Dai-2020}
\begin{equation}
\frac{ \partial \mathbf n_i}{\partial t}  
= - \mathbf n_i \times  \frac{\partial H_{\boldsymbol \lambda}} {\partial \mathbf n_i}  +  \mu \, \mathbf n_i \times (\mathbf n_i \times  
\frac{\partial H_{\boldsymbol \lambda} }{\partial \mathbf n_i} )
\label{eom2}
\end{equation}
with $\mu > 0$ the large$-t$ limit is a critical point
of $H_{\boldsymbol \lambda}$ since  (\ref{eom2}) gives
\begin{equation}
\frac{d H_{\boldsymbol \lambda}}{dt} = - \frac{\mu}{1+\mu^2}\sum\limits_{i=1}^N \left |  \frac{ \partial \mathbf n_i }{\partial t} \right |^2
\label{gilbert}
\end{equation}
so that the time evolution (\ref{eom2})
proceeds towards decreasing values of  $ H_{\boldsymbol \lambda}$ and the flow
continues until a critical point ($\mathbf n_{i}^\star \, , {\boldsymbol \lambda}^\star$) is reached.  
Whenever more than one solution exist, I  choose the one with smallest value of $H$.

\subsection{Simple examples}\label{ra_sec4}

\subsubsection{5.1 Triangular time crystal}

\begin{figure}
\centerline{\includegraphics[width=14.0cm]{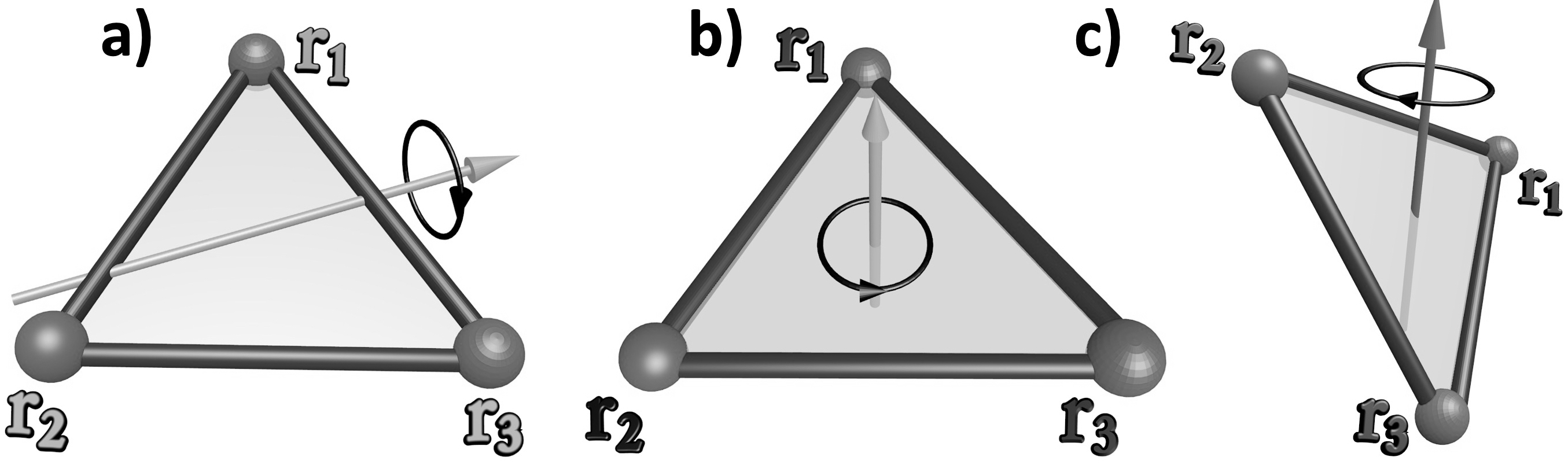}}
\caption{a): A timecrystalline triangle with $b=0$ and rotation axis determined by the $a_i$. b): A timecrystalline triangle with $a_i=0$, $b\not=0$ and 
the rotation axis is the symmetric normal to the triangular plane. c): For generic ($a_i,b$) the rotation axis points to generic direction. }
\label{fig-3}
\end{figure}

The first simple example is an equilateral triangle, with
Hamiltonian~\cite{Dai-2019} 
\begin{equation}
H = \sum\limits_{i=1}^3 a_i \mathbf n_i \cdot \mathbf n_{i+1} \ + \ b \mathbf n_1 \cdot (\mathbf n_2 \times \mathbf n_3)
\label{H3}
\end{equation}
Hamilton's equation (\ref{eom}) gives
\begin{equation}
\begin{split}
\partial_t \mathbf n_1 \ = \ & \mathbf n_1 \times (a_1 \mathbf n_2 + a_3 \mathbf n_3) + b \mathbf n_1 \times (\mathbf n_2 \times \mathbf n_3)  \\ 
\partial_t \mathbf n_2 \ = \ & \mathbf n_2 \times (a_2 \mathbf n_3 + a_1 \mathbf n_1) + b \mathbf n_2 \times (\mathbf n_3 \times \mathbf n_1)  \\
\partial_t \mathbf n_3 \ = \ & \mathbf n_3 \times (a_3 \mathbf n_1 + a_2 \mathbf n_2) + b \mathbf n_3 \times (\mathbf n_1 \times \mathbf n_2) 
\end{split}
\label{eom3d}
\end{equation}
By summing these equations I obtain
\[ 
\frac{d}{dt} ( \mathbf n_1 + \mathbf n_2 + \mathbf n_3 ) = 0
\]
This confirms that an initially closed chain remains closed for all $t$. Moreover, for general choice of parameters 
($a_i, b$) the {\it r.h.s.} of  (\ref{eom3d}) never vanishes; 
the minimum of $H$ for a closed chain is not a critical point of $H$; the minimum is
time dependent {\it i.e.} a time crystal. The time crystal describes rotation with an angular velocity
around  an axis, with direction  determined by  the parameters. See Figure \ref{fig-3}.   
%
%
%
%
%
%
%
%
%
%
%

\subsubsection{5.2 Knotted time crystals} 

\begin{figure}
\centerline{\includegraphics[width=14.0cm]{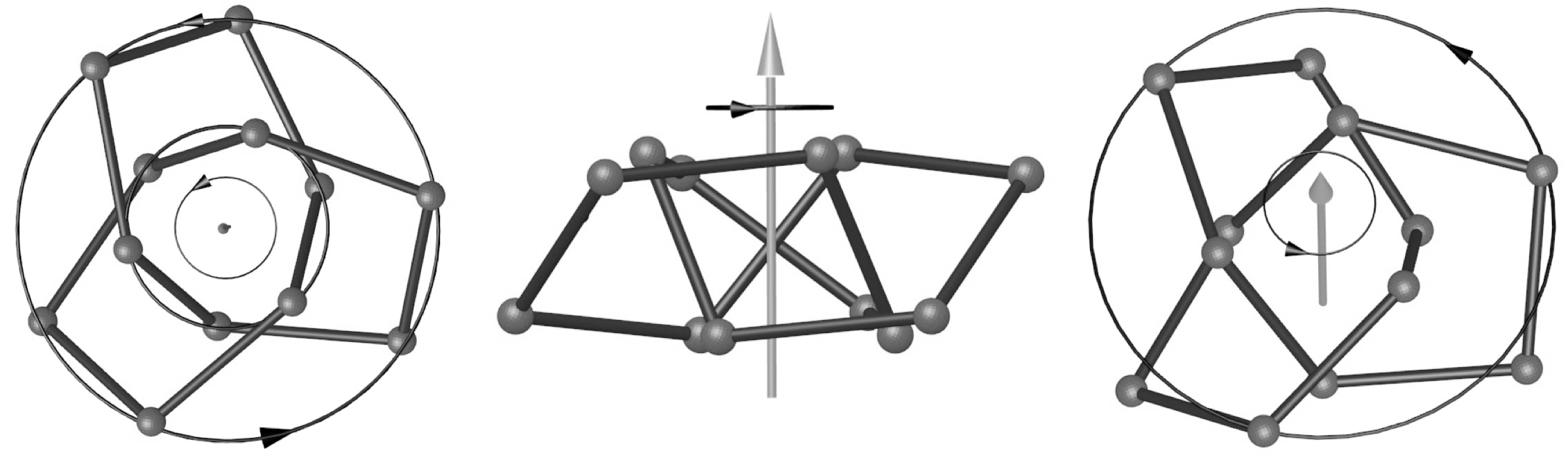}}
\caption{A topological time crystal, with the topology of a trefoil knot, viewed from three different directions. The energy function is given 
by (\ref{H12}), the rotation axis points in the direction of
$\boldsymbol \lambda^\star$ and the angular velocity is proportional to  $|\boldsymbol \lambda^\star|$.}
\label{fig-4}
\end{figure}

The second concrete, simple example involves only the long range energy function (\ref{U}), with Hamiltonian~\cite{Dai-2020} 
\begin{equation}
H = \frac{1}{2}  {\sum_{\buildrel{i,j=1}\over{i\not=j}}^{12} \left\{ 
\frac{1}{ |\mathbf r_i - \mathbf r_j| } \ + \  
\left( \frac{ 3/4 }{|\mathbf r_i - \mathbf r_j| }\right)^{12} \right\} }
\label{H12}
\end{equation}
The first term is a Coulomb repulsion between the vertices and the  second term is a short range Pauli repulsion that prevents chain crossing; 
in an actual molecule the covalent bonds can not cross each other.  The links connecting the $N=12$ vertices are chosen to
have the topology of a trefoil knot. The initial knot geometry is otherwise random. The Hamiltonian
is first minimized using the Landau-Lifschitz-Gilbert equation (\ref{eom2}).  The set-up is an example
of (\ref{eom}), (\ref{first}).   The Figure \ref{fig-4} shows the resulting minimum energy
time crystal, how it rotates according to (\ref{eom})
around the axis that coincides with the vector ${\boldsymbol \lambda}^\star$.
%
%
%
%
%
%
%
%
%
%
%

Various combinations of the local angles (\ref{H1234}) and the long distance interactions (\ref{U}) can be 
introduced. The ensuing energy functions can describe 
time crystalline structures, also with more elaborate knot topologies than a trefoil. 
At a higher level of realism,  the interaction centers at the vertices can be given their internal atomic structure.
Eventually, one ends up with a highly realistic all-atom molecular dynamics description of a polymer chain, which 
is the subject of the next Section.

%
%
%
%
%
%

\section{6. Cyclopropane and Sisyphus}\label{ra_sec4}

\subsection{6.1 Cyclopropane as a  time crystal}

All-atom molecular dynamics  simulations are the most realistic descriptions that are
presently  available to model 
chain-like  molecules. CHARMM
~\cite{charmm}
is an example of such a molecular dynamics  energy function,  and GROMACS~\cite{gromacs} is a user-friendly package for performing simulations.
 A typical molecular dynamics energy function contains the following terms; here the summations cover
all atoms of the molecule and can also extend {\it e.g.} over ambient water molecules.

\begin{equation}
\begin{split}
V(\mathbf x_i) &= \sum\limits_{bonds}k_{bi} ( l_i - l_{i0})^2 + \sum\limits_{angles} k_{ai} (\theta_i-\theta_{i0})^2 + \sum\limits_{torsions} V_i^n [ 1- \cos(n\omega_i - \gamma_i)]
\\
&+ {\sum_{{i\not=j}}} \left\{  \frac{ e_i e_j  } { |\mathbf x_{i} - \mathbf x_j | } +  \frac{\epsilon_{ij}}{2}
\left[   \left(\frac{ r_{min} } { |\mathbf x_{i} - \mathbf x_j | }\right)^{\!\! 12} - 2\left(\frac{ r_{min} } { |\mathbf x_{i} - \mathbf x_j | }\right)^{\!\! 6}\right] \right\}
\label{amber}
\end{split}
\end{equation}
The first term describes the stretching and shrinking of covalent bonds; it is not present in (\ref{H1234}), (\ref{U}) where
the Lie-Poisson bracket (\ref{n-bra}) preserves the bond lengths. 
The second and third terms account for the bending and twisting in the covalent bonds. These terms are akin 
quadratic/harmonic approximations to non-linear terms that are listed in (\ref{H1234}). The last term is a 
combination of the electromagnetic and Lennard-Jones interactions (\ref{U}).  
There can also be additional terms  such as the Urey-Bradley interaction between atoms separated by two bonds (1,3 interaction),
and improper dihedral terms for chirality and planarity; these can also be descrobed by the higher order terms in (\ref{H1234}).
The time evolution is always
computed from Newton's equation, with the force field that derives from (\ref{amber}).

In the case of a molecular chain,  a typical  characteristic  time scale for a covalent bond oscillation that is due to 
stretching and shrinking of the bond described by the first term in (\ref{amber}),  can be as short as 
a few femtoseconds.  In practical observations of molecular motions the  time scales are usually 
much  longer,  and the observed covalent bond lengths commonly correspond to time averaged values. 
For  the bending and twisting motions the characteristic time scales are much longer than for stretching and shrinking.
Thus  a separation of scales should take place so that an effective theory description becomes
practical. Indeed, many phenomena that are duly consequences of the free energy (\ref{amber}) can often be
adequately modeled  by an effective theory description. The effective theory energy function can be  a 
combination of terms such as those in (\ref{H1234}), (\ref{U}) and its dynamics can be 
described by the Lie-Poisson bracket (\ref{n-bra}), in a useful approximation.   

\begin{wrapfigure}{l}{0.4\textwidth}\vspace{-0pt}
 \begin{center}
    \includegraphics[width=0.3\textwidth]{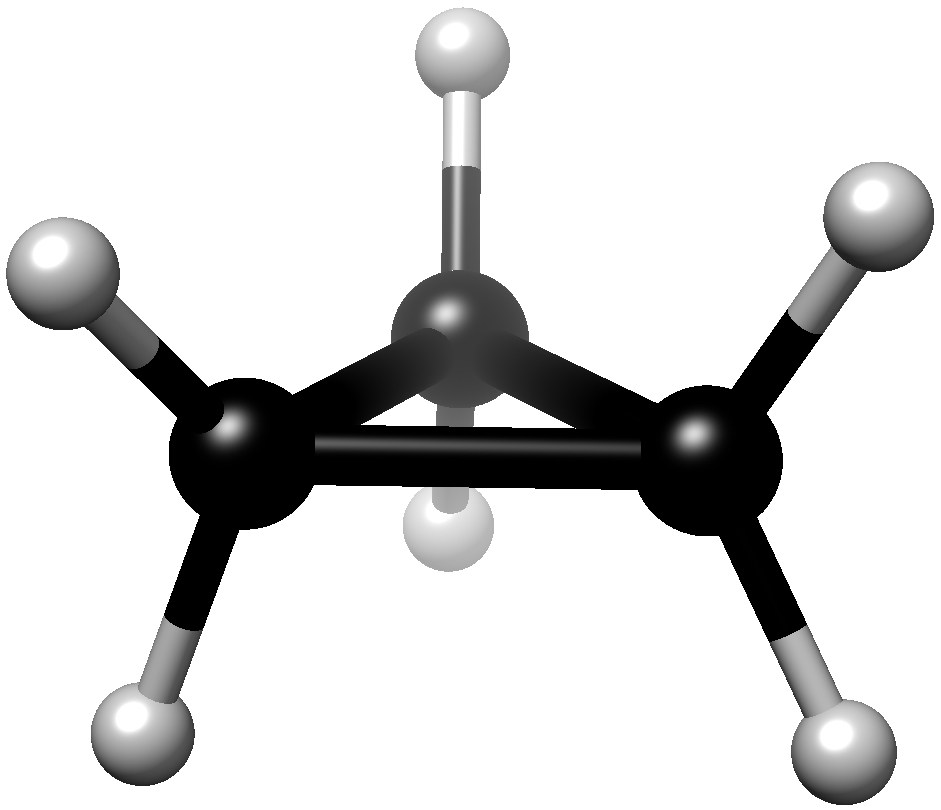}
      
      \vspace{-10pt}
  \caption{  \hspace{+3pt}\footnotesize  \sl \hspace{-0mm} Cyclopropane C$_3$H$_6$ with $D_{3h}$ molecular symmetry.}
  \end{center}
     \label{fig-5}
\end{wrapfigure}

To  investigate how the separation of scales takes place, 
and how an effective theory description emerges,  in [~\citenum{Peng-2021}]  all-atom molecular dynamics has been used 
to
simulate the ground state of  an isolated cyclopropane molecule C$_3$H$_6$ shown in Figure \ref{fig-5}. The force field
is CHARMM36m, in combination with GROMACS. The simulation starts with a search of the
%
%
%
%
%
%
%
%
%
%
minimum energy configuration, at a given ultra low internal temperature factor value; very low  temperature  
thermal oscillations could be  interpreted as mimicking quantum oscillations in a semiclassical description. 
The initial atomic coordinates can be found from the PubMed site ~\cite{PubMed} where the positions of 
the carbon and hydrogen atoms are specified with $10^{-14} \,$m precision. 
In [~\citenum{Peng-2021}]  the structure is further optimized so that it describes a local 
minimum of the CHARMM36m energy 
function, with full $D_{3h}$ molecular symmetry where the carbon-carbon bond angles are $60^{\mathrm o}$ and the hydrogen 
pairs are in full eclipse. 
Starting from this energy minimum all-atom trajectories are constructed, to simulate 10$\mu$s of cyclopropane dynamics, 
at different ultra low internal temperature factor values.  The simulations use double precision floating point  accuracy and 
the length of the simulation time step is 1.0$f$s; for the detailed set-up and for simulation details I refer to  [~\citenum{Peng-2021}]. 
All the individual atom coordinates  are followed and recorded,  and analysed
at different stroboscopic time steps $\Delta_s t$, during the entire 
10 $\mu$s molecular dynamics trajectory.

When the cyclopropane is simulated with the very low $\sim 0.067$K internal temperature factor value, 
and the all the atom positions are observed at every  $\Delta_s t= 100n$s  (or longer) stroboscopic time step, the molecule
rotates uniformly at constant angular velocity.  The axis of rotation  coincides 
with the (time averaged) center of mass axis that  is normal to the plane of the 
three carbon atoms. 
Remarkably, this stroboscopic rotational motion  is {\it identical} to the motion of a triangular Hamiltonian time crystal 
shown in Figure \ref{fig-3} b):  The dynamics is described by the generalized Landau-Lifschitz equation 
(\ref{eom}), with the Hamiltonian $H_2$ in (\ref{H1234}. The results confirm that the timecrystalline 
Hamiltonian $H_2$ is 
an effective theory for this stroboscopic motion, in the limit of long stroboscopic time steps.

\subsection{6.2 Spontaneous symmetry breaking}

Notably, the Lie-Poisson bracket  (\ref{n-bra}) breaks parity; if the sign on the {\it r.h.s.} is changed,   the rotation direction
in Figure {\ref{fig-3}) b) also changes. Obviously something similar needs to take place 
in cyclopropane for it to rotate in a particular direction:
The cyclopropane is {\it a priori} a highly symmetric molecule with $D_{3h}$  molecular symmetry; the carbon-carbon bond angles 
are all $60^{\rm o}$ and the dihedrals of all hydrogen pairs are fully eclipsed. In this maximally symmetric state there can not be any unidirectional rotational motion around the molecular symmetry axis that is normal to the plane of carbons, as there is no way to select between clockwise and counterclockwise rotational direction. 
However, the bond angles are much smaller than the optimum $109.5^{\rm o}$
angles of a normal tetrahedral carbon atom, and there is considerable 
torsional strain between the fully eclipsed hydrogen pairs~\cite{strain}. Thus one can expect 
that in the lowest energy ground state the $D_{3h}$  symmetry becomes spontaneously broken. By 
rigidity of covalent bond lengths I expect that the symmetry breakdown 
should be mainly due to a twisting of the dihedral angles, between the hydrogen pairs. 
This spontaneous symmetry breakdown selects a rotation direction, since parity is no longer a symmetry. 
The simulation results that I have described show that this indeed occurs: The unidirectional rotation 
around the triangular symmetry axis in the limit of long stroboscopic time steps is a manifestation of broken parity.

A simple  model free energy  can be introduced,  to demonstrate
how the spontaneous symmetry breaking due to strain in hydrogen pair dihedral angles
can take place. With $\theta_i$ ($i=1,2,3$) the dihedrals, the free energy is 
\begin{equation}
F(\theta_1,\theta_2,\theta_3) = \frac{1}{4}\sum\limits_{i=1}^3 g (\theta_i^2 - \alpha^2)^4  
\label{F1}
\end{equation}
The eclipsed configuration with all $\theta_i=0$ is a local maximum, and  a critical point of the free energy.
The minimum of (\ref{F1})
occurs when $\theta_i = \pm \alpha$, the value of $\alpha$ corresponds to the optimal value of the 
dihedral angle for  two staggered hydrogen pairs.  But in the cyclopropane molecule
the  three dihedrals are subject to the condition
\begin{equation}
\theta_1 + \theta_2 + \theta_3 = 0
\label{F2}
\end{equation}
Thus $\theta_i^{min} = \pm \alpha$ is not achievable for non-vanishing $\alpha$. 
Instead I need to find the minimum of (\ref{F1}) subject to the condition
(\ref{F2}). This is (again) a problem in constrained optimization, so I search for critical points of
\[
F_\lambda =  \frac{1}{4}\sum\limits_{i=1}^3 g (\theta_i^2 - \alpha^2)^4  + \lambda (\theta_1 + \theta_2 + \theta_3)
\]

%
%
%
%
%
 \begin{wrapfigure}{l}{0.4\textwidth}\vspace{-0pt}
 \begin{center}
    \includegraphics[width=0.4\textwidth]{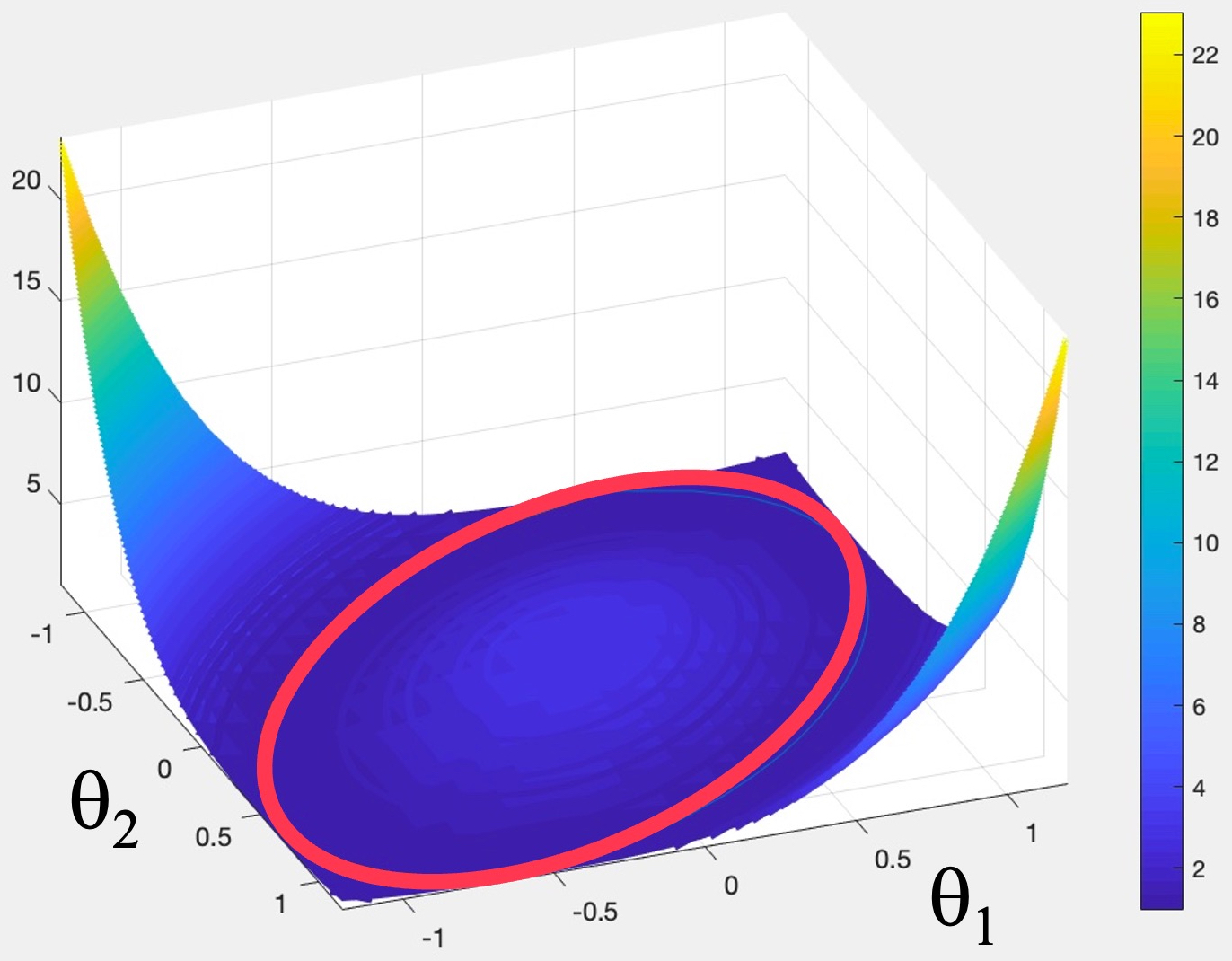}
      
      \vspace{-10pt}
  \caption{  \hspace{+3pt}\footnotesize  \sl \hspace{-0mm} Free energy (\ref{F1}) with $\theta_3$ given by
  (\ref{F2}), and its minimum.}
  \end{center}
     \vspace{-40pt}
     \label{fig-6}
\end{wrapfigure}
%
%
%
%
%
with $\lambda$ the Lagrange multiplier; I can rescale the angles, and set $a=g=1$ in which 
case the critical
points obey 
\begin{equation}
\theta_i (\theta_i^2-1) + \lambda = 0 
\label{F3}
\end{equation} 
in addition of (\ref{F2}).  I eliminate $\theta_3$ and plot the minimum of  the rescaled (\ref{F1}) as a function of 
$\theta_1$ and $\theta_2$. The result is shown in
Figure  6.  The minimum forms  a circular curve on the ($\theta_1,\theta_2$) plane around the origin. 
In particular,  the $D_{3h}$ symmetry becomes spontaneously broken, by the minimum. 
The Lagrange multiplier can be solved from (\ref{F3}). It is non-vanishing except when 
$\theta_1$ has the value  0 or $\pm 1$ and $\theta_2$ is $\pm 1$ or 0, respectively.  The non-vanishing of $\lambda$ is 
suggestive of a time crystal, in line with the general theory of Section 2.

\vspace{1.0cm}

\subsection{6.3 Sisyphus}

The Newtonian dynamics with the CHARMM36m force field  is much more complex that the generalized Landau-Lifschitz 
evolution (\ref{eom}) with the Hamiltonian $H_2$ in (\ref{H1234}). 
Thus  one can not expect that the cyclopropane molecule continues to display the same uniform, timecrystalline 
rotational motion 
when one decreases the length of the stroboscopic time step $\Delta_st$. The Figure \ref{fig-7} shows  how the rotational motion 
proceeds as function of the decreasing length of the stroboscopic time step; these Figures display 
the instantaneous stroboscopic value of the rotation angle $\theta(t)$ around the normal axis of the carbon triangle.

$\bullet~$ Figure  \ref{fig-7} a) shows the result for $\Delta_s t= 100n$s, when the effective theory description (\ref{eom}) with the Hamiltonian $H_2$ in (\ref{H1234})
is accurate: The rotation is uniform, with a constant angular velocity. 

$\bullet~$   In Figure  \ref{fig-7} b)  the length of the stroboscopic time step is decreased to 
$\Delta_s t = 20\,p$s. There is a very small amplitude ratcheting,  with an almost constant amplitude oscillations 
around the average value of the increasing rotation angle.  
 
$\bullet~$   In Figure \ref{fig-7} c)  the  stroboscopic time step is decreased to $\Delta_s t = 2.0 \,p$s. The amplitude of ratcheting 
oscillations around the average value of  $\theta$ has substantially increased:  The molecule turns around regularly 
and rotates in the opposite direction,  but at a slightly smaller relative value of
 angular velocity $\omega_s\equiv \dot \theta $. The period of ratcheting
 oscillations in $\theta(t)$ are also shorter than in Figure  \ref{fig-7} b).
 
$\bullet~$   Finally,  in Figure \ref{fig-7} d) the molecule is sampled with stroboscopic time steps $\Delta_s t = 200 \, f$s.  
Now the motion is becoming more chaotic, it consists of a superposition of 
very rapid back-and-forth rotations with different amplitudes and frequencies,  and with only a slow relative 
drift towards increasing 
values of $\theta(t)$.

The results shown in Figures \ref{fig-7} demonstrate how the separation of scales takes place in the simulated cyclopropane:
There is a continuous, smooth cross-over transition from a large-$\Delta_s t$ regime of  a uniform timecrystalline rotation,  through
an intermediate $\Delta_s t$ regime with increasingly ratcheting rotational motion, to a 
small-$\Delta_s t$ regime that is dominated by rapid back-and-forth oscillations, with different amplitudes and frequencies.
Remarkably, when the stroboscopic time step decreases,  the time evolution of the cyclopropane becomes qualitatively increasingly
similar to the Sisyphus dynamics reported in [~\citenum{sisyphus}]. Thus, the Sisyphus dynamics appears to provide a microscopic
level  explanation how timecrystalline  effective theory Hamiltonian dynamics emerges, at least in a small molecule such as
cyclopropane.

%
%
%
%
%
%
%
%
%
%
%
%
\begin{figure}
\centerline{\includegraphics[width=14.0cm]{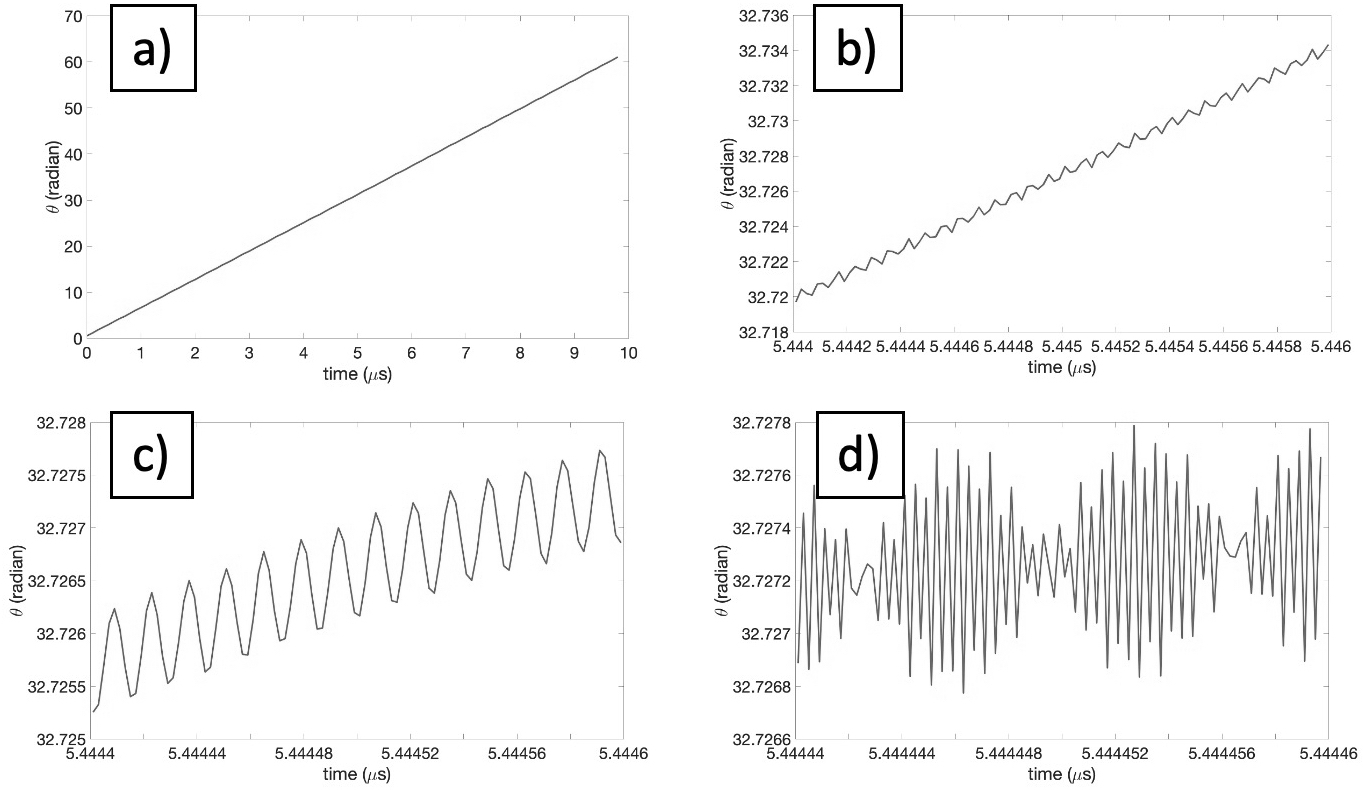}}
\caption{The evolution of the cyclopropane rotation angle during a $10 \,\mu$s CHARMM36m simulation and 
with $T=0.067$K internal temperature factor value, recorded 
with decreasing stroboscopic time steps: a) The regime of uniform rotation, here with $\Delta_s t= 100\, n$s. b) and c)
 describe ratcheting regime, with  $\Delta_s t = 20\,p$s in b) and 
$\Delta_s t = 2.0 \,p$s in c). Finally d) with $\Delta_s t = 200\,f$s is in the regime dominated by Sisyphus dynamics. }
\label{fig-7}
\end{figure}
%
%
%
%
%
\vskip 0.4cm

\section{7. Rotation without angular momentum}

Newton's equation with  the CHARMM36m all-atom molecular dynamics force field 
preserves the angular momentum, and angular momentum is also well preserved in a GROMACS 
numerical simulation. Since the initial cyclopropane has no 
angular momentum, the rapid back-and-forth oscillations and in particular the uniform timecrystalline rotational motion
that is observed,  are in apparent violation of angular momentum conservation. 
The resolution of the paradox is that the cyclopropane is not a rigid body. It is a deformable body, and 
a deformable body with at least three movable 
components can rotate simply by changing its shape~\cite{Guichardet-1984,Shapere-1989b}. 
A falling cat is a good example, 
how it can maneuver and rotate in air to land on its feet.

In the case of a cyclopropane molecule,  when the internal temperature factor has a non-vanishing
value, the covalent bonds oscillate so that the shape of the molecule continuously 
changes; in an actual molecule there are also quantum mechanical zero-point oscillations.  Such shape changes are
minuscule, but over a long trajectory their effects can accumulate and self-organize into an apparent 
rotational motion. This is what is described in the Figures \ref{fig-7}. 

The analysis of results in [~\citenum{Peng-2021}] confirm that the angular momentum of the simulated cyclopropane is
conserved, and vanishes  with numerical precision
during the entire $10 \, \mu$s simulation trajectory.  For this, one evaluates 
the accumulation of infinitesimal rotations, with each rotation 
corresponding to that of an instantaneous rigid body. An infinitesimal 
rigid body rotation can be defined using {\it e.g.} Eckart frames, in terms of the instantaneous positions and velocities of all the carbon and hydrogen
atoms around the center of mass. In our simulations 
these are recorded at every $\Delta \tau = 10^{-15} \, f$s time step $n$ during 
the entire $10 \, \mu$s production run. 
The instantaneous values $L(n)$  of the corresponding rigid body angular momentum component
along the normal to the instantaneous plane of the three carbon atoms can then be evaluated, together 
with the corresponding instantaneous moment of inertia values $ I(n)$.  
This gives  the following instantaneous ``rigid body'' angular velocity values 
\[
\omega(n) = L(n) / I(n)
\] 
When these are summed up the result is the accumulated ``rigid body'' rotation angle, at each simulation step $n$:
\begin{equation}
\vartheta(n) = \omega(n) \Delta \tau + \vartheta(n-1) \ = \  \sum\limits_{i=1}^{n} \frac{L(i)}{I(i)} \Delta \tau  
\label{vartheta}
\end{equation}
In full compliance with the conservation of angular momentum and the vanishing of its initial value,
it is found~\cite{Peng-2021} that in all the  $10 \,\mu$s production run simulations the accumulated values (\ref{vartheta}) always 
remains less than $\sim 10^{-6}$ radians, 
for all $n$, and the Figure \ref{fig-8} shows a typical example: There is no observable net rotation of the cyclopropane molecule
due to ``rigid body'' angular momentum, with numerical precision.    
Accordingly any systematic rotational motion that exceeds $\sim 10^{-6}$ radians during a production run 
simulation must be emergent,  and  entirely due to shape deformations. 
%
%
%
%
%
%
\begin{figure}
\centerline{\includegraphics[width=11.0cm]{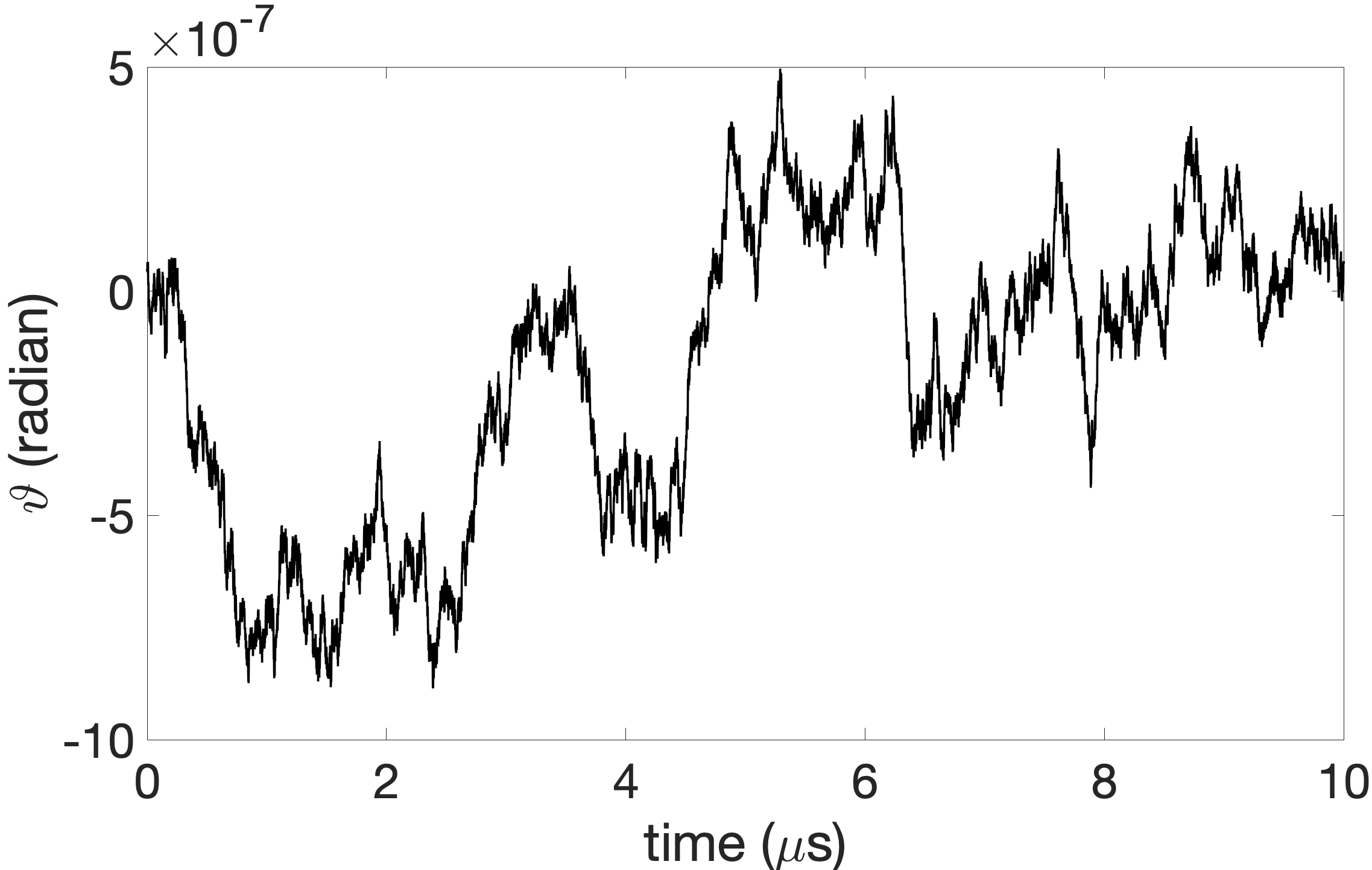}}
\caption{ The evolution of the rigid body rotation angle $\vartheta(n)$ in (\ref{vartheta}), during  a typical cyclopropane simulation. }
\label{fig-8}
\end{figure}
%
%
%
%
%

The original articles on rotation by shape deformations are 
[~\citenum{Guichardet-1984,Shapere-1989b}]. Reviews can be found in [~\citenum{Littlejohn,Dai-2020}] and
here I follow  [~\citenum{Dai-2020}]:
Consider the (time) 
$t$-evolution of three equal mass point particles that form 
the corners $\mathbf r_i$  ($i=1,2,3$) of a triangle. I assume that there are 
no external forces so that the center of mass does not move, 
\begin{equation}
\mathbf r_1 (t) +  \mathbf r_2 (t) + \mathbf r_3 (t) = 0
\label{r123}
\end{equation}
for all t.  I also assume that the total angular momentum  vanishes,
\begin{equation}
\mathbf L \ = \ \mathbf r_1 \times \dot {\mathbf r}_1 + \mathbf r_2 
\times \dot {\mathbf r}_2 + \mathbf r_3 \times \dot {\mathbf r}_3 \ = \ 0
\label{Lz}
\end{equation}

I now show that nevertheless,  the triangle can rotate by changing its shape.  To describe this 
rotational motion, I place the triangle to the $z=0$ plane and with the center of mass at the origin
($x,y$)=($0,0$).   
Two triangles then have the same shape if they only differ by a rigid rotation on the plane, around the $z$-axis.  I can describe 
shape changes by shape coordinates  $\mathbf s_i(t)$ that I assign to each vertex of the triangle. They describe  
all possible triangular shapes, in an unambiguous fashion and in particular with no extrinsic rotational motion, when I
demand that the vertex $\mathbf s_1(t)$ always lies on the positive $x$-axis with $s_{1x}(t) >0$ and $s_{1y}(t)=0$,  
and the vertex $\mathbf s_2(t)$  has $s_{2y}(t)  >0$ but $s_{2x}(t) $ can be arbitrary.
The coordinates $\mathbf s_3(t)$ of the third vertex are then fully determined by the
demand that the center of mass remains  at the origin:
\[
\mathbf s_3(t) =  - \mathbf s_1(t) - \mathbf s_2(t)
\] 

Now, let the shape of the triangle change arbitrarily, but in a $\mathrm T$-periodic fashion. As a consequence
the  $\mathbf s_i(t)$ evolve also in a
$\mathrm T$-periodic fashion,
\[
\mathbf s_i(t+T) = \mathbf s_i(t)
\] 
as the triangular shape traces a closed loop 
$\Gamma$ in the space of all possible triangular 
shapes.  

 At each time $t$ the shape coordinates  $\mathbf s_i(t)$ and  the 
space coordinates  $\mathbf r_i(t) $ are related by a rotation around the $z=0$ plane,
\begin{equation}
\left( \begin{array}{c} r_{ix}(t) \\ r_{iy}(t) \end{array} \right) = 
\left( \begin{array}{cc}  \cos \theta(t)  &  - \sin \theta(t) \\  \sin \theta(t)  & \  \cos \theta(t) \end{array} \right) \left( \begin{array}{c} s_{ix}(t) \\ s_{iy}(t) \end{array} \right) 
\label{theta}
\end{equation}
Initially $\theta(0)=0$, but if there is any net rotation due to shape changes we have $\theta(\mathrm T)\not= 0$ 
so that the triangle has rotated during the period,  by an angle  $\theta(\mathrm T)$. I
substitute (\ref{theta}) into (\ref{Lz}) and I get  
\begin{equation}
\theta(\mathrm T)  \ \equiv \ \int\limits_0^{\mathrm T} \! dt \, \frac{d\theta(t)}{dt} \,  
\ = \ \int\limits_0^{\mathrm T} \! dt \, \, \frac{
\sum\limits_{i=1}^{3}  \left\{ s_{iy} \dot s_{ix}  - s_{ix} \dot s_{iy}\right\}  
}
{ 
\sum\limits_{i=1}^3 \mathbf s^2_i
}
\label{dotheta}
\end{equation}
and in general this does not need to vanish, as I show in the next sub-section

\section{8. Towards timecrystalline universality}
I now evaluate (\ref{dotheta}) in the case of a  time dependent 
triangular structure, with three point-like interaction centers at the corners. The
shape  changes are externally driven so that the shape coordinates evolve as follows: 
\begin{equation}
\mathbf s_1 (t) \ = \  \frac{1}{\sqrt{3}} \left( \begin{matrix} \cos [ a  \sin \omega_1 t ] 
 \\  \hspace{0.cm} 0   \end{matrix}    \right) \ \  \ \ \& \ \  \ \ \mathbf s_2(t) 
\ = \  \frac{1}{\sqrt{3}} \left( \begin{matrix} \cos [  a \sin  \omega_2t  + \frac{2\pi}{3} ] \\   \hspace{0.cm} \sin ( 
\frac{2\pi}{3} ) \end{matrix} \right)
\label{rot1}
\end{equation}
I  choose $a=0.1$ and $\omega_2=2\omega_1=3$, substitute in (\ref{dotheta}) 
and evaluate the time integral numerically.  The Figures \ref{fig-9} a)-d)  show how 
the angle $\theta(t)$ evolves, when observed with different stroboscopic time steps.

When I compare the  Figures \ref{fig-7} and \ref{fig-9} I observe a striking qualitative similarity:
Except for the scales, the corresponding panels are almost identical. 
%
%
%
%
%
%
\begin{figure}
\centerline{\includegraphics[width=14.0cm]{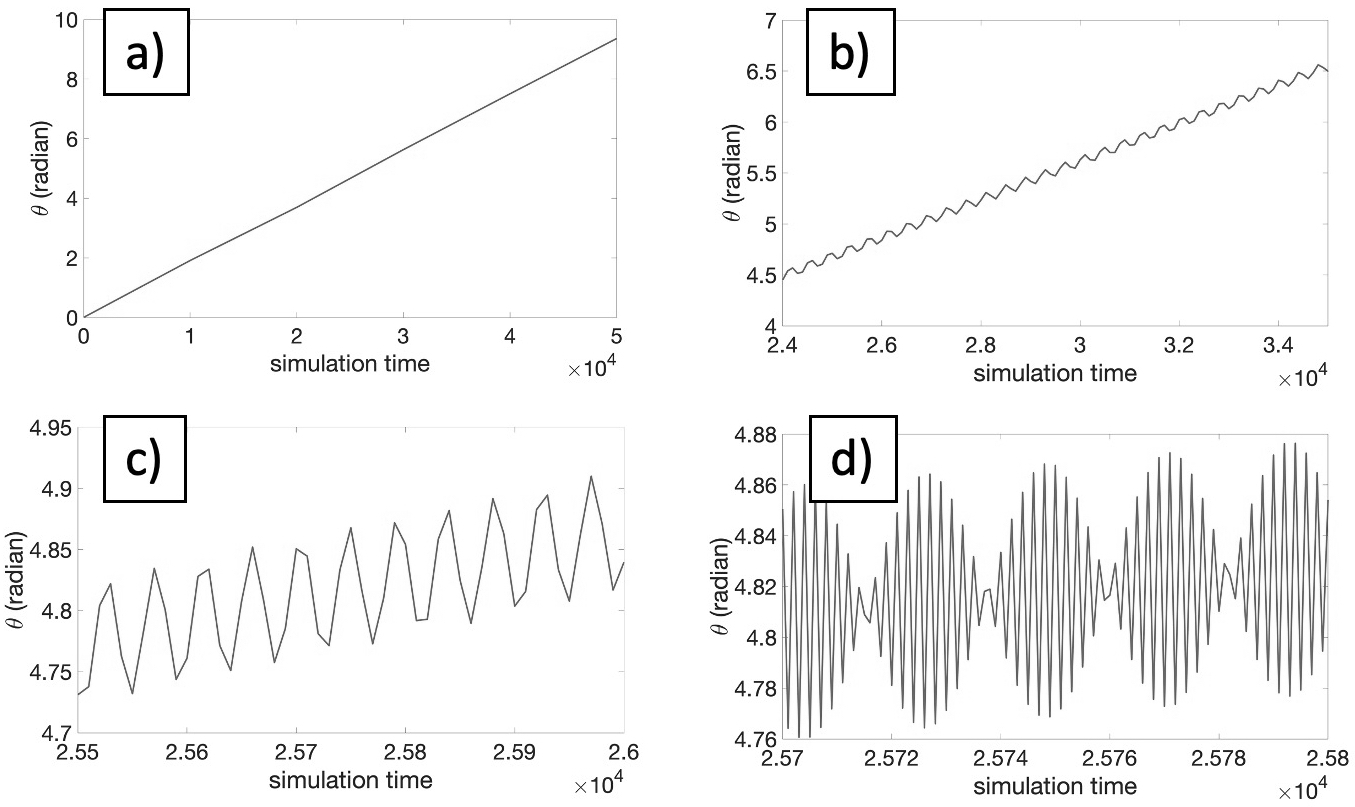}}
\caption{ The evolution of the angle $\theta(t)$  of (\ref{dotheta}) for the shape changes (\ref{rot1}),
and with stroboscopic time steps. In panel a) $\Delta_s t  = 10^3$, In panel b) $\Delta_s t  = 10^2$,
In panel c) $\Delta_s t = 10$, and in panel d) $\Delta_s t  = 1$.}
\label{fig-9}
\end{figure}
%
%
%
%
%
In particular, even though the shape changes (\ref{rot1}) are externally driven, at large stroboscopic time scales
the time evolution again coincides with that of the autonomous Hamiltonian time crystal in Figure \ref{fig-3} b).

The strong qualitative similarity between results in Figure \ref{fig-7} and \ref{fig-9} proposes that the 
Sisyphus dynamics of [~\citenum{sisyphus}] and the ensuing ratcheting at larger stroboscopic time scales, 
is akin a universal route, how an oscillatory short time scale 
stroboscopic evolution becomes self-organized into a Hamiltonian time crystal, at large 
stroboscopic time scale.  

Finally, when I expand the integrand (\ref{dotheta}), (\ref{rot1}) in powers of the amplitude $a$, 
the result is
\[
\frac{ d\theta }{dt} \ = \ - \frac{1}{2} \omega_2 a \cos\omega_2 t  
+ \frac{\sqrt{3}}{12} a^2 \left\{ \omega_2 \sin 2\omega_2 t - \omega_1 \sin 2\omega_1 t \right\} - \frac{1}{8} \omega_2 a^3 \cos 
3\omega_2 t 
\]
\begin{equation}
+ \frac{1}{16} \omega_2 a^3 \left\{ \cos(\omega_2 + 2\omega_1)t + \cos(\omega_2 - 2\omega_1)t \right\} 
+ {\mathcal O}(a^4)
\label{small-a}
\end{equation}
Thus, exactly when $\omega_2 = \pm 2 \omega_1$ will the large time limit of the 
rotation angle $\theta(t)$ increase linearly in time as follows,
\begin{equation}
\theta(t) \ \buildrel {{\rm large}-t} \over {\longrightarrow}  \ \frac{1}{16} \, \omega_2 a^3 t
\label{unirot}
\end{equation}
in line with Figure \ref{fig-9} a). But when $\omega_2 \not= \pm 2 \omega_1$ the large-$T$  limit of the integral 
(\ref{dotheta}) vanishes, by Riemann's lemma. Similar high sensitivity of the rotation angle is also observed in
the case of the cyclopropane, where the role of the parameter $a$ is played by the internal temperature factor of
the molecule.

\section{Concluding remarks} 
The concept of a time crystal has made a  long journey.  It started as 
a beautiful idea that was soon ridiculed as a fantasy. From there it has progressed
to the frontline of theoretical physics research, with high expectations for remarkable future applications. But the
notion is  still very much under development. The conceptual principles are still under debate and remain to be
finalized, there are  several  parallel and alternative  lines of research to follow.  
The present article describes only my own personal way, how I try  to understand what is a time crystal.
My interest in the subject started from discussions with Frank, and  I am  only able to describe what I have 
learned myself, by doing things on my own, with Frank, and with my close colleagues.  There is undoubtedly 
much that I have not covered, but I leave it  to  others who know things better.

\section{Acknowledgements} 
I thank Anton Alekseev, Jin Dai, Julien Garaud, Xubiao Peng and Frank Wilczek for collaboration on various aspects of
the original work described here.  My research is supported by the Carl Trygger Foundation 
Grant CTS 18:276
and by the Swedish Research Council under Contract No. 2018-04411 and by Grant 
No. 0657-2020-0015 of the Ministry of Science and Higher Education of Russia.  I also 
acknowledge COST Action CA17139.

\section{Personal recollections} 

I conclude with a short  personal recollection of the remarkable way how Betsy and Frank have ended up, 
to spend part of their time in Sweden and  Stockholm University.

I first met Betsy and Frank personally in Aspen, during the summer 1984 session. Incidentally,  we overlapped there 
with Michael Green and John Schwarz, during the time when they gave birth to the modern string theory. At  that time
Frank and I shared a more  modest interest, and we discussed certain ideas around the SU(3) Skyrme model. 
Unfortunately my self-confidence at the time was not at a level, to meet his challenges. 

After Aspen we met several times,  including Santa Barbara, Cambridge, Uppsala, Tours, Stockholm and elsewhere,
not to forget  their lovely
summer house in New Hampshire. Our discussions were always very enjoyable.
Frank even  included me in his official delegation to attend the events during the Nobel Week,  when he received the 2004 Physics
Prize.  

In 2007, around the time when Nordita was moving to Stockholm, Frank told me that he had plans  for a
sabbatical.  I succeeded in attracting him to spend half a year in Sweden, jointly between Nordita and my  Department
at Uppsala University. The other half of his sabbatical he spent at Oxford University, where I also visited him. There, he told me about his 
dream to walk across the Great Britain. Apparently he hatched the first, very early  
ideas of a time crystal during this walk. I now regret
I did not ask to join him.  

While in Uppsala,   Frank and I got the idea to organize a Nobel Symposium in graphene. 
The Symposium took place in Summer 2010, only a few months before the discovery of 
graphene was awarded the Physics Nobel Prize: Ours was the first ever Nobel Symposium
that took place the same year  the Prize was awarded on the subject of a Symposium.  
I understand that the many great talks, walks and discussions during the Symposium helped 
to  decide who got the Prize.  Two years later, I hosted Frank when he became honorary doctor at
Uppsala University.  A little later Frank came back to  Uppsala, this time to collect 
Nobel chocolates that he won from a bet  with Janet Conrad, on the discovery of the Higgs boson.

Both Betsy and Frank seemed very happy with their sabbatical stay, and with all other experiences they had at 
Stockholm and Uppsala. So I started to talk with Frank about the idea, that they could spend  a little 
longer time  in Sweden, on a regular basis.  In particular, I told Frank that the Swedish Research Council  
had occasional calls on a  funding program for International Recruitments, with very generous conditions.  
In 2014 the opportunity raised:  Soon after the Nobel Symposium on topological insulators at H\"ogberga where
Frank gave the summary talk,  I learned that the call was again being opened, and that   
this was probably the last call in the program. I contacted Frank, now with more determination.  
He was at least lukewarm, so I proposed my colleagues at Uppsala University that we should make an 
application to try and recruit Frank. 
Initially, I received several very positive, in fact some enthusiastic, replies from my colleagues at the 
Department of Physics and Astronomy.   But then came one strongly negative reply, 
from Maxim Zabzine who at the time was responsible for  the administration of theoretical physics. Maxim stated that
he did not want Uppsala University to make any effort, whatsoever, to apply for a Research Council  grant for Frank. 

I did not see any point to try and change Maxim's strong opinion, in particular since the deadline for the application
was approaching.
 Instead, I immediately contacted 
Lars Bergstr\"om at Stockholm University. I asked him 
if Stockholm University would be willing to submit an application to the Research Council, on Frank's behalf.
I reckon that I contacted Lars only two days before the Nobel Prize of Physics 2014 was announced.
Lars was at the time the Permanent Secretary of the Nobel  Physics Committee 
and for sure he had his hands full in preparing the announcement. However, already the next day I received  
a reply from Anders Karlhede, then Vice President of Stockholm 
University. Anders  wrote that my proposal had been discussed with the President of Stockholm University Astrid S\"oderberg Widding, 
and that Stockholm University is delighted  to submit  an application to the  Research Council.

Soon after  the Nobel Prize in Physics 2014 was announced and Lars became more free, he and
Anders started to work on the formal application; I understand they were also joined by Katherine Freese 
who was then  Director of Nordita. At that time I  were in 
China with Frank.  Frank  was on his first-ever trip to China, and I coordinated the visits in Beijing together with Vincent Liu. 
From Beijing we all continued  to Hangzhou  for  the memorable inaugural  events of Wilczek Quantum Center at Zhejiang Institute of 
Technology. 

During our travel in China I had several long discussions with Frank about the Research Council  grant application. He had his doubts, 
and I made my best to persuade him. I remember vividly the decisive discussion: After a breakfast at our hotel, 
a historic place that used to be Mao Zedong's favorite retreat at the West Lake in Hangzhou, 
Frank and I were sitting together in the lobby.  In Stockholm, the Research Council application was ready to be 
submitted, the deadline was only a couple of days away. But
Stockholm University still needed Frank to sign a formal letter of interest for the application, and Frank was
hesitant. However,  after some lengthy discussion I got his  signature, and I sent the signed letter right away to Anders.
In two months time we received a decision from the Research Council that Stockholm University's application for  the 
International Recruitement Grant for Frank had been approved.  

However,  it was too early to call it a home run. For that,
I still needed some advice from Anders B\'ar\'any, who is the creator of Nobel Museum. Anders 
came up with the brilliant idea, of a curator position for Betsy at Nobel Museum. To launch this,
I teamed up with G\'erard Mourou who invited Frank and Betsy, and a delegation from Nobel Museum,
to the Symposium on Fresnel, Art and Light at Louvre in Paris. Frank gave a beautiful talk, 
and Betsy made contacts with Louvre art curators.  With a strong support from Astrid and Anders things worked 
out impressively  for Betsy at the Nobel Museum, where her achievements included
the extraordinary Feynman Exhibition at ArtScience Museum of 
Marina Bay in Singapore 2018-2019 that she largely organized with Frank's support and great help from K K Phua.  
With some additional aid and support from here and there,  including Grand hospitality by Katherine
Freese  at a right time in summer 2015, and in particular with  the consistent and very strong support from Astrid and Anders, 
Betsy and Frank were finally convinced to try and start their  present journey  in Stockholm.
I am really grateful that I can follow and share so much of it with them.

%


\vspace{-0mm}
\begin{thebibliography}{100}
\bibitem{wilczek-2012} F. Wilczek, Phys. Rev. Lett. {\bf 109} 160401 (2012)
\vspace{0mm}
\bibitem{shapere-2012b} A. Shapere,  F. Wilczek,  Phys. Rev. Lett. {\bf 109} 160402 (2012)
\bibitem{bruno-2013} P. Bruno,  Phys. Rev. Lett. {\bf 111}  070402 (2013)
\vspace{0mm}
\bibitem{watabane-2015} H. Watanabe, M. Oshikawa, Phys. Rev. Lett. {\bf 114} 251603 (2015)
\vspace{0mm}
\bibitem{yamamoto-2015} N. Yamamoto, Phys. Rev. {\bf D92}  085011 (2015)
\vspace{0mm}
\bibitem{Sacha-2015} K. Sacha, Phys. Rev. {\bf A91} 033617 (2015)
\vspace{0mm}
\bibitem{Sacha-2016}  K. Sacha, D. Delande, Phys. Rev. {\bf A94} 023633 (2016)
\vspace{0mm}
\bibitem{khemani-2016} V. Khemani, A. Lazarides, R. Moessner, S.L. Sondhi, Phys. Rev. Lett. {\bf 116} 250401 (2016)
\vspace{0mm}
\bibitem{else-2016a} D.V. Else, C. Nayak, Phys. Rev. {\bf B93} 201103 (2016)
\vspace{0mm}
\bibitem{else-2016} D.V. Else, B. Bauer, C. Nayak  Phys. Rev. Lett. {\bf 117} 090402 (2016) 
\vspace{0mm}
\bibitem{else-2016b} D. V. Else, B. Bauer, C. Nayak, Phys. Rev. {\bf X7}  011026 (2017)
\vspace{0mm}
\bibitem{Yao-2017} N. Y. Yao, A. C. Potter, I.-D. Potirniche, A. Vishwanath, Phys. Rev. Lett. {\bf 118}  030401 (2017)
\vspace{0mm}
\bibitem{zhang-2017} J. Zhang {\it et.al.} Nature {\bf 543} 217 (2017) 
\vspace{0mm}
\bibitem{lukin-2017} S. Choi {\it et.al.} Nature {\bf 543} 221 (2017) 
\vspace{0mm}
\bibitem{Rovny-2018} J. Rovny, R. L. Blum,  S. E. Barrett, Phys. Rev. Lett. {\bf 120} 180603 (2018). 
\vspace{0mm}
\bibitem{Dai-2019} J. Dai, A.J. Niemi, X. Peng, F. Wilczek Phys. Rev.  {\bf A99} 023425-9 (2019)
\vspace{0mm}
\bibitem{anton} A. Alekseev, J. Dai, A.J. Niemi  JHEP {\bf 08} 035 (2020) 
\vspace{0mm}
\bibitem{Dai-2020} X. Peng, J. Dai, A.J. Niemi,  New J. Phys. {\bf 22}  085006 (2020)
\vspace{0mm}
\bibitem{Fletcher-1987} R.~Fletcher, \emph{{Practical Methods of Optimization}}. (Wiley, Chichester New York, 1987)
\bibitem{Nocedal-1999} J.~Nocedal and S.~J. Wright, \emph{{Numerical Optimization}}, Springer Series
  in Operations Research. (Springer, Heidelberg, 1999)
\vspace{0mm}
\bibitem{Marsden} J.E. Marsden and T.S. Ratiu, {\it Introduction to Mechanics and Symmetry A Basic Exposition of 
Classical Mechanical Systems} Second Edition (Springer Verlag, New York, 1999)
\vspace{0mm}
\bibitem{sisyphus} A. D. Shapere, F. Wilczek Proc. Natl. Acad. Sci. U.S.A. {\bf 116} 18772  (2019)
\vspace{0mm}
\bibitem{Guichardet-1984} A. Guichardet,  
 Ann.  l'Inst. Henri Poincar\'e     {\bf 40}  329 (1984)
\vspace{0mm}
\bibitem{Shapere-1989b}A. Shapere,   F. Wilczek,  
Am. J. Phys.    {\bf 57} 514  (1989)
\vspace{0mm}
\bibitem{Littlejohn} RG Littlejohn, RG Reinsch, 
Rev.  Mod. Phys.   {\bf 69} 213 (1997)
\vspace{0mm}
\bibitem{sonin} E. B. Sonin, {\it Dynamics of Quantised Vortices in Superfluids}  (Cambridge University Press, Cambridge, 2016).
\vspace{0mm}
\bibitem{Garaud-2021a} J. Garaud, J. Dai, A.J. Niemi, JHEP {\bf 2021} 7 (2021)
\vspace{0mm}
\bibitem{Wilczek-1982a} F. Wilczek, Phys.Rev.Lett. {\bf  48} 1144 (1982)
\vspace{0mm}
\bibitem{Wilczek-1982b} F. Wilczek, Phys.Rev.Lett. {\bf 49} 957 (1982)
\vspace{0mm}
\bibitem{Milne} L. M. Milne-Thomson, {\it Theoretical Hydrodynamics} 5th edition  (Dover, New York, 1996)
\vspace{0mm}
\bibitem{charmm} J Huang et.al., Nature Meth. {\bf 14} 71 (2017)
\vspace{0mm}
\bibitem{gromacs} {\tt http://manual.gromacs.org/}
\vspace{0mm}
\bibitem{Peng-2021} X. Peng, J. Dai, A.J. Niemi,  New J. Phys. {\bf 23}  073024 (2021)
\vspace{0mm}
\bibitem{PubMed} {\tt https://pubchem.ncbi.nlm.nih.gov/compound/6351}
\vspace{0mm}
\bibitem{strain}  V. Dragojlovic, 
Chem. Texts {\bf 1} 14 (2015) 
\vspace{0mm}

 \end{thebibliography}
\end{document}